\documentclass[aps,pra,twocolumn,amsmath,amssymb,superscriptaddress,10pt,nobibnotes]{revtex4-2}
\usepackage{preamble}

\begin{document}
	

\title{Maximum information states for coherent scattering measurements}
\author{Dorian Bouchet}
\altaffiliation{\href{mailto:dorian.bouchet@univ-grenoble-alpes.fr}{dorian.bouchet@univ-grenoble-alpes.fr}}
\affiliation{Nanophotonics, Debye Institute for Nanomaterials Science and Center for Extreme Matter and Emergent Phenomena, Utrecht University, P.O. Box 80000, 3508 TA Utrecht, Netherlands}
\affiliation{Present address: Université Grenoble Alpes, CNRS, LIPhy, 3800 Grenoble, France}
\author{Stefan Rotter}
\affiliation{Institute for Theoretical Physics, Vienna University of Technology (TU Wien), 1040 Vienna, Austria}
\author{Allard P.\ Mosk}
\affiliation{Nanophotonics, Debye Institute for Nanomaterials Science and Center for Extreme Matter and Emergent Phenomena, Utrecht University, P.O. Box 80000, 3508 TA Utrecht, Netherlands}


\begin{abstract}
The use of coherent light for precision measurements has been a key driving force for numerous research directions, ranging from biomedical optics~\cite{park_quantitative_2018,taylor_interferometric_2019} to semiconductor manufacturing~\cite{osten_optical_2012}. Recent work demonstrates that the precision of such measurements can be significantly improved by tailoring the spatial profile of light fields used for estimating an observable system parameter~\cite{van_putten_nonimaging_2012,shechtman_optimal_2014,balzarotti_nanometer_2017,ambichl_super-_2017,yuan_detecting_2019,juffmann_local_2020,bouchet_influence_2020}. 
These advances naturally raise the intriguing question of which states of light can provide the ultimate measurement precision~\cite{giovannetti_advances_2011}. 
Here, we introduce a general approach to determine the optimal coherent states of light for estimating any given parameter, regardless of the complexity of the system. 
Our analysis reveals that the light fields delivering the ultimate measurement precision are eigenstates of a Hermitian operator which quantifies the Fisher information based on the system's scattering matrix~\cite{rotter_light_2017}. To illustrate this concept, we experimentally show that these maximum information states can probe the phase or the position of an object that is hidden by a disordered medium with a precision improved by an order of magnitude as compared to unoptimized states. Our results enable optimally precise measurements in arbitrarily complex systems, thus establishing a new benchmark for metrology and imaging applications~\cite{osten_optical_2012,barrett_foundations_2013}.
\end{abstract}

\maketitle

	
No physical observable can be determined with absolute certainty. 
Instead, the noise inherent in any measurement process sets a fundamental limit on the precision that a physical observable can be estimated with~\cite{kay_fundamentals_1993,giovannetti_advances_2011}. 
Whenever light or other kinds of electromagnetic radiation are involved in a measurement, a necessary condition to reach this ultimate precision is to optimize the spatial distribution of the radiation field in the measured system~\cite{barrett_foundations_2013}. To achieve this goal, a crucial task is to identify the spatial pattern that should be imprinted on the incoming field in order to get the maximum information out of it. First progress in this direction has recently been made using wavefront shaping techniques and metasurfaces to precisely estimate lateral displacements~\cite{van_putten_nonimaging_2012,yuan_detecting_2019}, fluorophore positions~\cite{shechtman_optimal_2014,balzarotti_nanometer_2017}, spectral shifts~\cite{ambichl_super-_2017} or phase variations~\cite{juffmann_local_2020}. 

A central challenge that remains unresolved, however, is to identify a unifying approach to reach the ultimate precision limit that is applicable even to complex scattering systems. Earlier work suggests that such an approach should be connected to the concept of Fisher infor\-mation~\cite{kay_fundamentals_1993,giovannetti_advances_2011,shechtman_optimal_2014,bouchet_influence_2020}, which quantifies the amount of information relevant to the estimation of a given parameter from measured data. 
However, for the generic case of a complex medium, the Fisher information is intrinsically linked to the microstructure of the medium~\cite{bouchet_influence_2020}, which is not only overwhelmingly complex in realistic systems but also typically unknown. 

Here, we overcome this difficulty by expressing the Fisher information in terms of a Hermitian operator that depends on the system's optical scattering matrix.
Based on this idea, we introduce and experimentally demonstrate a direct approach to generate optimal coherent states of light for parameter estimation. These light states are shown to be specifically tailored even to a complex system, not only with respect to the specific observable of interest, but also with respect to the position of the observer. By unambiguously identifying these optimal light states, we establish a new general benchmark for metrology and imaging applications~\cite{osten_optical_2012,barrett_foundations_2013}. 
Furthermore, in the ideal case for which all optical modes supported by the system are accessible to the observer, our analysis reveals that maximum information states are, at the same time, the optimal states for optical micro-manipulation~\cite{ambichl_focusing_2017,horodynski_optimal_2020}, thereby uncovering a fundamental relationship between information theory and measurement backaction.

To set up this approach we recall that a measurement scheme is optimal when the measurements, the estimation function, and the choice of the incident state are all optimal concurrently~\cite{giovannetti_advances_2011}. To realize this situation for coherent states of light, we start with a general model of scattering measurements on a complex medium parameterized by a scalar parameter $\theta$ (Fig.~1). This parameter can be a global parameter characterizing the entire scattering medium. It can also be a local parameter of limited spatial extent such as the phase or the position of a small phase object hidden behind a scattering material, as in our experiments. We illuminate the medium from the far field by an incident coherent state $|E^\mathrm{in}\rangle$ characterized by the coefficients $\{E_1^\mathrm{in},\dots,E_M^\mathrm{in}\}$ in $M$ spatial modes, which are individually addressed using wavefront shaping techniques~\cite{mosk_controlling_2012}. The far field of the outgoing coherent state $|E^\mathrm{out}\rangle$  is then characterized by the coefficients $\{E_1^\mathrm{out},\dots,E_N^\mathrm{out}\}$ in $N$ spatial modes using a homodyne detection scheme, which introduces an external reference beam and measures the resulting number of photons in each spatial mode. The number of outgoing spatial modes can be taken as low as $N=1$, a feature that can be relevant for applications requiring a fast single-channel detector.

\begin{figure}
	\begin{center}
		\includegraphics[width=\linewidth]{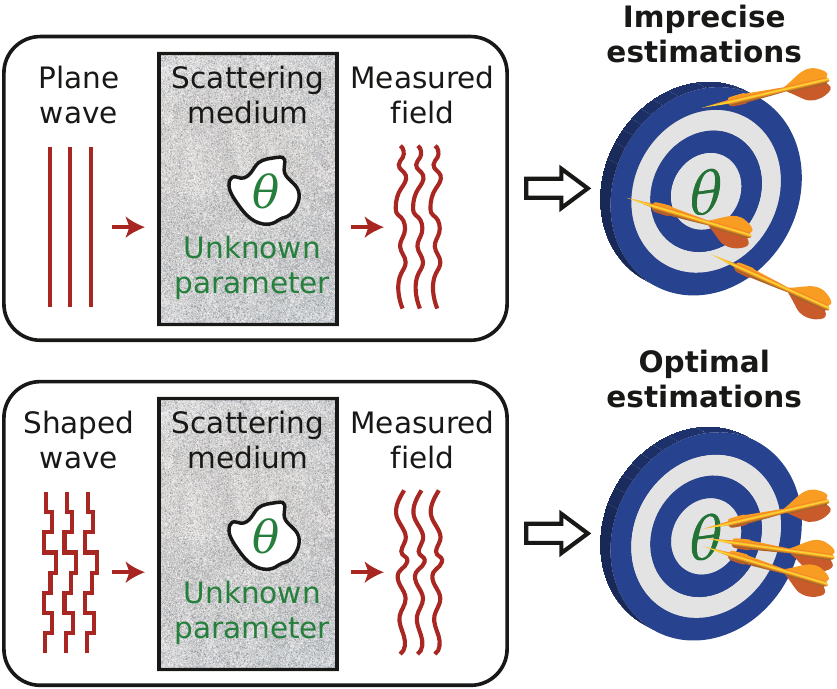}
	\end{center}
	\caption{\textbf{Fig. 1 | Principle of an optimal coherent scattering measurement.} 
	A scattering medium is characterized by an unknown parameter $\theta$. This parameter is estimated by illuminating the medium with coherent light and by measuring the outgoing field state via a homodyne detection scheme. In many cases, plane-wave illumination leads to imprecise estimations (top panel). The optimal incident states we generate here using wavefront shaping techniques enable optimal estimations (bottom panel).}
\end{figure}

Noise fluctuations in the measured data fundamentally limit the achievable precision on the determination of $\theta$. This limit is mathematically expressed by the Cramér-Rao inequality, which sets a lower bound on the variance of unbiased estimators of $\theta$. In general, the Cramér-Rao bound is given by the reciprocal of the Fisher information $\mathcal{J}(\theta)= \mathbb{E} \, ([ \partial_\theta \ln p(X;\theta) ]^2) $ where $X$ is a $N$-dimensional random variable representing the data, $p(X;\theta) $ is a joint probability density function and $\mathbb{E} $ denotes the expectation operator acting over noise fluctuations~\cite{kay_fundamentals_1993}. Here, we assume that this noise arises only from the quantum fluctuations of coherent states, and not from other possible noise sources such as sample-to-sample or wave-to-wave fluctuations~\cite{fang_concentration--measure_2018}. Considering that noise fluctuations are statistically independent for any two different outgoing modes, we derived the following simplified expression for the Fisher information (see Supplementary Information~S1.1):
\begin{equation}
\mathcal{J}(\theta)= 4 \sum_{k=1}^{N} \left\vert\partial_\theta E^\mathrm{out}_k\right\vert^2 \, .
\label{eq_fisher_gaussian}
\end{equation}
This expression is obtained by considering a specific homodyne detection scheme, composed of a high-intensity reference beam that interferes with the $k$-th spatial mode with a phase $\phi_k = \arg (\partial_\theta E_k^\mathrm{out})$.  Alternatively, one can also calculate the quantum Fisher information $\mathcal{I}(\theta)$, which sets a more general bound on the variance of unbiased estimators holding for any positive-operator-valued measure~\cite{braunstein_generalized_1996}. When evaluating the quantum Fisher information for an $N$-mode coherent state composed of separable pure states, we find that it coincides with Eq.~\eqref{eq_fisher_gaussian} so that $\mathcal{I}(\theta)=\mathcal{J}(\theta)$ (see Supplementary Information~S1.2), thereby demonstrating that the homodyne detection scheme considered here is optimal for the estimation of $\theta$.

We consider an incident state $|E^\mathrm{in}\rangle$ as being optimal for the Fisher information it generates when the corresponding outgoing state $|E^\mathrm{out}\rangle$ maximizes Eq.~\eqref{eq_fisher_gaussian} for a given number of incident photons. In order to identify this state, the pivotal quantity is the scattering matrix $S$ of the medium~\cite{rotter_light_2017}, which relates incoming and outgoing states via $ |E^\mathrm{out}\rangle =S |E^\mathrm{in}\rangle $. Using this relationship, the Fisher information can be written in the quadratic form $\mathcal{J}(\theta)=4 \langle  E^\mathrm{in} | F_\theta | E^\mathrm{in} \rangle $, where we used bra-ket notations for the complex inner product. We designate the term $F_\theta$ in the center of this expression as the {\it Fisher information operator} that takes on the remarkably simple form (see Supplementary Information~S1.3)
\begin{equation}
F_\theta= (\partial_\theta S) ^\dagger \partial_\theta S \; ,
\end{equation}
where $\dagger$ stands for the conjugate transpose. Since this operator $F_\theta$ is Hermitian already by construction, the incident state that maximizes the Fisher information is given by the eigenstate associated with the largest eigenvalue of $F_\theta$. Furthermore, in the limit of small parameter variations, we obtained a closed-form expression of the minimum variance unbiased estimator (see Supplementary Information~S1.4), which is the optimal estimation function. Importantly, the variance of this estimator always reaches the Cramér-Rao bound, even for small numbers of incident photons. This confirms that the Fisher information is here the relevant quantity to assess the precision achievable with different light states. 

\begin{figure*}[t]
	\begin{center}
		\includegraphics[width=16.5cm]{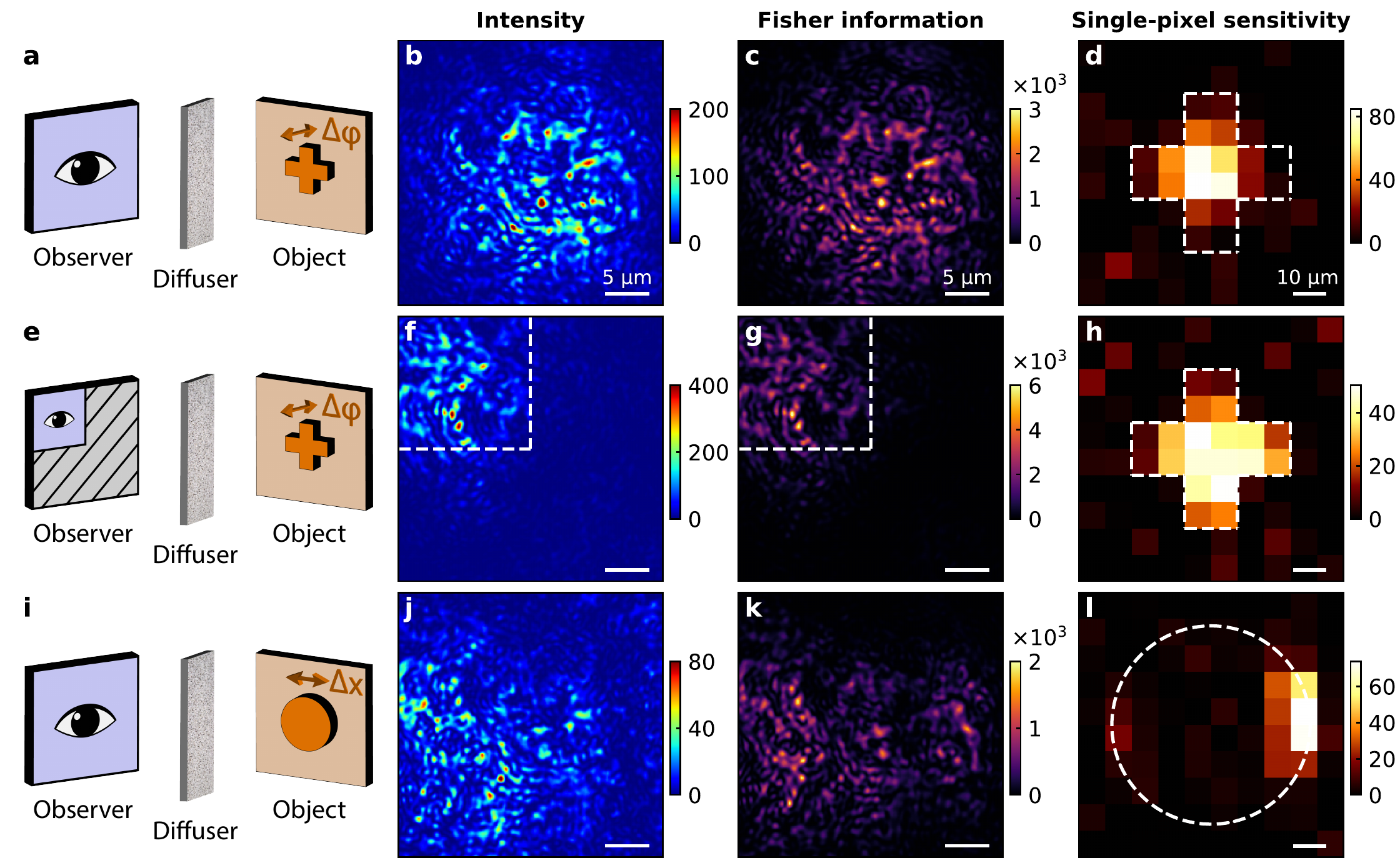}
	\end{center}
	\caption{\textbf{Fig. 2 | Characteristics of maximum information states.}
	\textbf{a}, Sketch of the experiment: the observer (left) is a camera with a field of view covering $880$\,\textmu m$^2$, separated by a diffuser (middle) from a cross-shaped target object (right) that induces a phase shift $\varphi$ as our observable parameter of interest. \textbf{b}, \textbf{c},~Measured spatial distributions of the intensity and of the Fisher information per unit area for the optimal state with maximum overall information content, respectively normalized by the average signal intensity and by the average Fisher information per unit area under plane-wave illumination. \textbf{d},~Single-pixel sensitivity measured by shifting the phase of each pixel in the target area of the hidden SLM for the optimal state, normalized by the average single-pixel sensitivity under plane-wave illumination. The cross-shaped target is indicated by white dashed lines. \textbf{e}--\textbf{h},~Analogous to \textbf{a}--\textbf{d} when the field of view of the camera covers a reduced area of $220$\,\textmu m$^2$, as delimited by white dashed lines in \textbf{f} and \textbf{g}. The maximum information state fully adjusts to the changes in the observer by delivering the Fisher information here primarily to the limited field of view. \textbf{i}--\textbf{l},~Analogous to \textbf{a}--\textbf{d} when the observable parameter is the horizontal position $x$ of a circular phase object. Also here the maximum information state adapts to the change in observable by redirecting its incoming intensity to those pixels on the very right or left edge of the target object that are most strongly affected by a lateral displacement $\Delta x$.}
\end{figure*}

In our optical experiments (wavelength $532$\,nm), we first choose, as the observable parameter $\theta$ that we aim to estimate, the phase shift $\varphi$ generated by a small cross (total length $48$\,\textmu m) displayed by a spatial light modulator (SLM), as represented in Fig.~2a. This target is hidden $1.2$\,mm behind a ground glass diffuser (scattering angle $15^\circ$). Our experimental setup includes a second SLM used to control the phase of the incident field and a detection scheme based on off-axis holography (see Extended Data Fig.~1). The experimental procedure starts by measuring three reflection matrices for different values of $\varphi$, which allows us to access both the reflection matrix $r$ and its derivative $\partial_\varphi r$ for $2437$ incident states and $2465$ outgoing states (see Methods). Even though the measured reflection matrix is large, it constitutes only a sub-part of the full $S$-matrix of the medium. This is not a limitation of our approach, which also applies to non-unitary and non-square matrices. Defining the operator $f_\varphi=(\partial_\varphi r)^\dagger \partial_\varphi r$, the eigenvector associated with the largest eigenvalue of $f_\varphi$ is the optimal incident state based on the available knowledge. Illuminating the medium with this state using the input SLM, we measured the spatial distributions of the outgoing signal intensity (Fig.~2b) and of the Fisher information per unit area (Fig.~2c), respectively normalized by the average signal intensity and by the average Fisher information per unit area under plane-wave illumination. Averaging over the field of view, the maximum information state generates a $300$-fold enhancement of the Fisher information along with a $20$-fold intensity enhancement, as compared to the average values measured under plane-wave illumination.

We then checked explicitly in which way the maximum information state is shaped in the near-field of the cross-shaped phase perturbation. To this end, we measured the single-pixel sensitivity, which is defined here as the total Fisher information in the outgoing state for phase variations on each individual pixel on the hidden SLM. These measurements are performed by successively varying the phase shift induced by individual pixels sequentially instead of varying the phase shift induced by the cross-shaped target as a whole (see Methods). We find that the maximum information state is primarily sensitive to a few pixels in the center of the cross (Fig.~2d), which confirms its economical wave function design. We emphasize that optimal states are not conceived to reveal the shape of the target, but to estimate the phase shift it induces. This does not require the intensity to be uniformly distributed on the area defining the target, explaining why its shape is not always fully revealed.

An important characteristic of maximum information states is their specificity with respect to both the position of the observer and to the observable of interest. To explicitly probe the influence of the observer, we repeat the experiment from above using, however, a reduced field of view in the detection plane (Fig.~2e--h and Extended Data Fig.~2). Remarkably, the maximum information state readjusts to this new observer by redirecting the spatial distributions of its outgoing signal intensity (Fig.~2f) and of its Fisher information per unit area (Fig.~2g) straight to the selected observer area. This confirms the essential role played by the set of optical modes we incorporate into the definition of maximum information states. While the single-pixel sensitivity (Fig.~2h) is similar to what is observed when the whole field of view is taken into account (compare with Fig.~2d), it is here more uniformly distributed inside the cross-shaped area. Indeed, such a redistribution in the plane of the hidden SLM is associated with smaller diffraction angles, thus allowing for a reduced spatial extent of the intensity distribution in the plane of the diffuser. To demonstrate also the specificity of maximum information states with respect to the observable of interest, we repeat the experimental procedure by choosing the horizontal position $x$ of a circular phase object (radius $30$\,\textmu m) as being the observable parameter of interest (Fig.~2i--l and Extended Data Fig.~3). The single-pixel sensitivity is now localized on the right or left edge of the object, which attests that maximum information states selectively focus light waves onto those specific areas of the object that have the most pronounced dependence on the observable of interest. Interestingly, the maximum information state typically focuses on a single edge rather than on both edges simultaneously, which is to be expected as the mirror symmetry in the system is broken by the diffuser.

To exemplify the remarkable advantages these features provide for measurements with very few photons where the estimation precision is limited by shot noise, we reduce the incident photon flux by placing a neutral density filter (fractional transmittance $8.3 \times 10^{-7}$) in the optical path. Under these conditions, illuminating the medium with a plane wave does not allow reliable estimations for the phase shift induced by target. This is confirmed by calculating the precision limit $\sigma_{\textsc{crb}}$ for the $2437$ plane waves used to construct the reflection matrix (see Supplementary Information~S2.1). For these plane waves, the median of the obtained phase-distribution is $1.8$\,rad, with a minimum value of $0.60$\,rad (Fig.~3a). In contrast, the precision limit associated with the maximum information state equals $0.066$\,rad, an entire order of magnitude smaller than the minimum value measured for plane waves. While reaching this precision limit would require amplitude and phase modulation of the incident field, we only modulate its phase by the input SLM in the experiments. We find that the precision limit associated with phase-only modulation of the maximum information state equals $0.078$\,rad, a value that is only slightly larger than that for a joint amplitude and phase modulation. This observation corroborates that our approach is also very robust with respect to small errors or imperfections in the preparation of the incident state. 

\begin{figure}[t]
	\begin{center}
		\includegraphics[width=\linewidth]{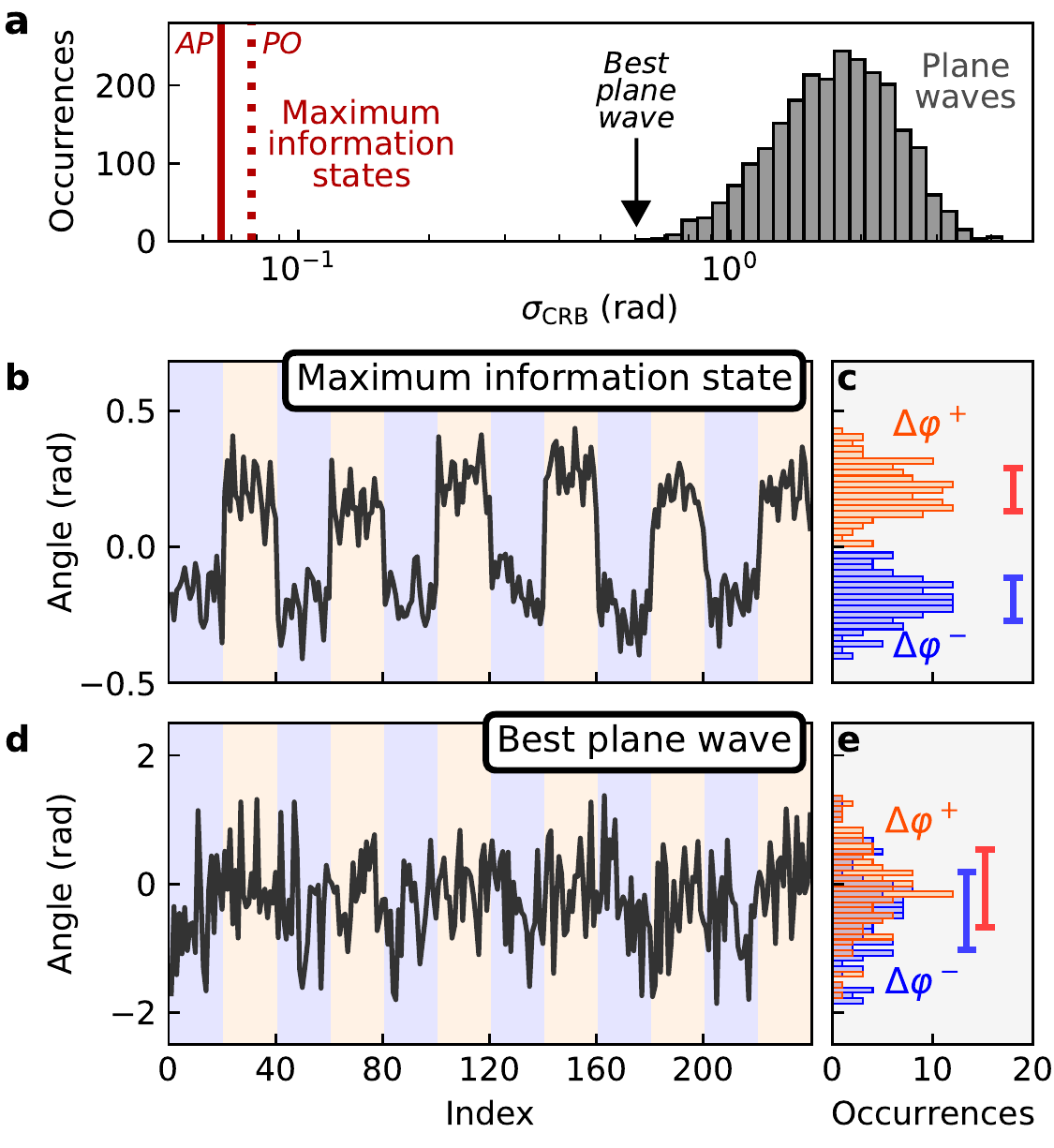}
	\end{center}
	\caption{\textbf{Fig. 3 | Demonstration of optimal estimations.}
	\textbf{a},~Histogram of precision limits for the $2437$ plane waves used to construct the reflection matrix (the observable parameter is here the phase shift $\varphi$ of the cross-shaped target shown in Fig.~2a). As compared to the best plane wave, the precision limit is improved by an order of magnitude with maximum information states (AP, amplitude and phase modulation; PO, phase-only modulation). \textbf{b},~Estimated phase shift of the cross-shaped target as a function of measurement index for measurements performed by illuminating the medium with the maximum information state. \textbf{c},~Histograms of estimated angles for a positive phase shift ($\Delta \varphi^+=+0.25$\,rad) and a negative phase shift ($\Delta \varphi^-=-0.25$\,rad) applied by the hidden SLM. The length of error bars equals $2 \, \sigma_{\textsc{crb}}$. \textbf{d}, \textbf{e},~Analogous to \textbf{b} and \textbf{c} for measurements performed by illuminating the medium with the best plane wave.}
\end{figure}

In order to demonstrate the practical implementation of the estimation process with a maximum information state, we use this state to illuminate the medium and perform a sequence of measurements. The observable parameter we estimate is the phase shift $\varphi$ induced by the cross-shaped target, which is varied every $20$ measurements between $-0.25$\,rad and $0.25$\,rad in a step-like manner. From the knowledge of the expected outgoing state and its derivative with respect to $\varphi$, we can construct the minimum variance unbiased estimator of $\varphi$, applicable to small parameter variations (see Supplementary Information~S2.2). We can then estimate the value of $\varphi$ from measurements of the outgoing state (see Extended Data Fig.~4). Even though only approximately $24,000$ incident photons are probing the medium per measurement, each step can be clearly resolved (Fig.~3b,c). The observed standard error on the estimates is $0.11$\,rad, which corresponds to a transverse displacement of the target of $9.6$\,nm (see also Extended Data Fig.~5 for estimations of lateral displacements). Importantly, the observed standard error on the estimates almost reaches the precision limit predicted by the Cramér-Rao inequality. This confirms that shot noise is the dominant source of noise in our experiment, and demonstrates that the precision limit is correctly predicted from reflection matrix measurements. We applied the same procedure for measurements performed by illuminating the medium with the best plane wave used to construct the reflection matrix. The precision limit is then much larger than the step size, thereby prohibiting a clear detection of the phase steps (Fig.~3d,e).

The Fisher information operator $F_\theta$ employed in our analysis not only constitutes an operational tool for the identification of maximum information states but it is also deeply connected to fundamental concepts in optics. Provided that reciprocity holds in terms of the transposition symmetry of the scattering matrix ($S^\mathsf{T}=S$), we can show (see Supplementary Information~S1.5) that the iterative phase conjugation of a small perturbation $\Delta \theta$ converges towards the largest eigenvalues of $F_\theta$. This insight not only suggests a potentially useful approach to identify maximum information states but also provides a new understanding of existing focusing procedures based on time-reversed adapted perturbation~\cite{zhou_focusing_2014,ma_time-reversed_2014,ruan_focusing_2017}. Moreover, in the ideal case of a unitary $S$-matrix ($S^\dagger=S
^{-1}$) where all optical modes supported by the system are accessible, we obtain the identity $F_\theta = Q_\theta^2$, with $Q_\theta=-i S^{-1} \partial_\theta S$ being the generalized Wigner-Smith operator. This operator was recently introduced to design optimal light fields for optical micro-manipulation in complex media~\cite{ambichl_focusing_2017,horodynski_optimal_2020}. The simple relation between $F_\theta$ and $Q_\theta$ suggests a new interpretation of $Q_\theta$ as the operator representing the measurement backaction on the conjugate quantity to $\theta$. The eigenstates of $Q_\theta$ (also called {\it principal modes}) have the remarkable property of being insensitive with respect to small variations in $\theta$ except for a global phase factor~\cite{carpenter_observation_2015,xiong_spatiotemporal_2016}. Likewise, the Fisher information of maximum information states is exclusively enclosed in variations of the global phase of the outgoing state, rather than in the state's intensity variations or speckle decorrelation. Nevertheless, this property strictly holds only for a unitary $S$-matrix and deviations from this property are observed in our experiments (see Supplementary Information~S3.3). It is also interesting to discuss the special case for which the observable of interest is the dielectric constant $\epsilon$ of a target. Indeed, provided that $S^\dagger=S^{-1}$, eigenstates of $Q_\epsilon$ maximise the integrated intensity inside the target~\cite{horodynski_optimal_2020}, which implies that maximum information states are then, at the same time, the light states that maximize power delivery to the target. Finally, while we have considered a scalar parameter $\theta$ in our analysis, our formalism also enables the identification of light states that maximize the trace of the multi-parameter Fisher information matrix (see Supplementary Information~S1.6), thus providing a possible strategy to perform precise estimations of multiple parameters.

To summarize, we demonstrate, for the first time, a method to identify and produce coherent light fields that are optimal for precision measurements in a complex environment. This work opens up new perspectives to enhance the performance of imaging techniques~\cite{barrett_foundations_2013} and, by simultaneously engineering the Fisher information operator itself, could also be used to improve the sensitivity of existing nano-structured sensing devices~\cite{hodaei_enhanced_2017,chen_exceptional_2017}.
We emphasize that our results are generally applicable to any parametric dependence of a wave field and can thus be transferred also to other types of waves such as in acoustics~\cite{gerardin_full_2014} or in the microwave regime~\cite{shi_transmission_2012}. 
Finally, it can be expected that the bound derived in this work for coherent states will be surpassed using quantum metrology protocols~\cite{giovannetti_quantum_2006,giovannetti_advances_2011,fiderer_maximal_2019}. In this context, our results provide a general benchmark to assess the performance of quantum states optimized for parameter estimation, and suggest a new path towards the identification of optimally-sensitive quantum states of light using scattering matrices of complex systems.




\section*{Acknowledgements}
The authors thank M. van Beurden, J. Bosch, S. Faez, M. Horodynski, M. K{\"u}hmayer, P. Pai, and J. Seifert for insightful discussions, and P. Jurrius, D. Killian and C. de Kok for technical support. This work was supported by the Netherlands Organization for Scientific Research NWO (Vici 68047618 and Perspective P16-08) and by the Austrian Science Fund (FWF) under project number P32300 (WAVELAND).


\section*{Methods}

\paragraph*{Optical setup}

The light source is a continuous wave solid-state laser (Coherent OBIS 532-120 LS FP) emitting at $532$\,nm. The laser light is coupled to a single-mode polarization-maintaining fiber and out-coupled using a collimator (Schäfter-Kirchhoff, 60FC-L-4-M75-01). The beam is separated into a signal path and a reference path using a 50:50 beamsplitter. In the signal path, the light beam passes a 50:50 beamsplitter and is then modulated by the input SLM (Holoeye Pluto VIS). It can then pass different neutral density filters which are mechanically placed or removed. The incident power on the objective is: $36.1$\,\textmu W when all density filters are removed from the optical path; $275$\,nW when a neutral density filter of optical density $2$ (ND2) is placed in the optical path (measured fractional transmittance $\mathcal{T}'=0.76 \times 10^{-2}$); $30$\,pW when a neutral density filter of optical density $6$ (ND6) is placed in the optical path (measured fractional transmittance $\mathcal{T}=0.83 \times 10^{-6}$).

The surface of the SLM is imaged onto a ground glass diffuser using a 4f system composed by a $200$\,mm lens and a $\times50$ objective (Nikon 50X CFI60 TU Plan Epi ELWD, 0.6 NA). A 90:10 beamsplitter is located between the lens and the objective. The diffuser is made by polishing a microscope coverslip. The resulting scattering angle, defined from the full width at half maximum of the transmitted intensity distribution, is approximately $15^\circ$. The diffuser is mounted on the windows of the hidden SLM (Holoeye Pluto BB) at a distance of $1.2$\,mm. The surface of the diffuser is then imaged using a 4f system composed by the objective and a $200$\,mm lens using a CCD camera (AVT Stingray F145-B) with an exposure time of $300$\,\textmu s. Camera acquisition is triggered by the input SLM in order to limit phase noise due to the flicker of the SLM. Before reaching the camera, the beam passes a polarizer to ensure that only the horizontal component of the field is measured, and also passes a 90:10 beamsplitter to recombine the reference path with the signal path. Both quadratures of the complex field are measured in a single shot using digital off-axis holography. Using a tilted beam instead of the optimally-shaped reference field does not impact the shape of the incident field optimized for the estimation of $\varphi$, but only leads to a reduction in the Fisher information by a factor of two (see Supplementary Information~S2.1).

\paragraph*{Construction of the reflection matrix and its derivative.}

We construct densely-sampled reflection matrices relating incident field states to reflected ones~\cite{popoff_measuring_2010,pai_optical_2020}. Reflection matrix measurements are performed with no density filter in the signal path. To illuminate the medium, we vary the incidence angle of a Gaussian beam characterized by a full-width at half maximum of $120$\,\textmu m, which is $4$ times larger than the field of view (such Gaussian beams are refereed to as plane waves in the manuscript). We probe $M=2437$ different incidence angles, covering a numerical aperture of $0.5$. The sampling is performed using a triangular lattice (in Fourier space), with a lattice constant taken to be the smallest angular separation at which the complex inner product of nearest-neighbor fields drops to zero.

For each incident angle, we record the reflected field using digital off-axis holography~\cite{takeda_fourier-transform_1982,cuche_spatial_2000}. The method relies on a reference beam which is tilted by an angle with respect to the reflected signal beam. The complex field is then numerically reconstructed from the measured holograms by digitally filtering the zero-order component. The reflected field is sampled using a triangular lattice, with a lattice constant that we determine by finding the distance between the maximum and the first minimum of the autocorrelation of a random field. We use $N=2465$~different sampling points, covering an area of $880$\,\textmu m$^2$ on the surface of the diffuser. The reflection matrix $r$ is therefore constructed column by column and, as a result, we obtain a $2465\times2437$ matrix. This matrix can then be used to faithfully predict the outgoing state for any incident state (see Supplementary Information~S3.1).

In the initial experiment, the parameter of interest is the phase shift $\varphi$ generated by a cross-shaped target object. The reflection matrix $r$ is then measured at an angle $\varphi=\varphi_0$ by setting the phase shift induced by all pixels of the hidden SLM to a given value. Note that $\varphi_0$ can be set to zero without loss of generality. In order to estimate the derivative of the reflection matrix with respect to $\varphi$, we measure two other reflection matrices $r(\varphi_0-\Delta \varphi)$ and $r(\varphi_0+\Delta \varphi)$, where $\Delta \varphi=0.54$\,rad. We can then estimate $\partial_\varphi r$ by applying the centered finite difference scheme $\partial_\varphi r \simeq [r(\varphi_0+\Delta \varphi)-r(\varphi_0-\Delta \varphi)]/(2\Delta \varphi) $. This matrix can then be used to faithfully predict the derivative of the outgoing state with respect to $\varphi$, for any incident state (see Supplementary Information~S3.2).

In another experiment, the parameter of interest is chosen to be the lateral position $x$ of a circular object. In this case, a super-Gaussian function of order $7$ is displayed by the hidden SLM, with a full-with at half maximum equal to $60$\,\textmu m. The phase difference between the object and the background is set to $\pi/2$\,rad and the reflection matrix $r$ is then measured at a position $x=x_0$. In order to estimate the derivative of the reflection matrix with respect to $x$, we measure two other reflection matrices $r(x_0-\Delta x)$ and $r(x_0+\Delta x)$, where $\Delta x=5$\,\textmu m. We can then estimate $\partial_x r$ by applying the centered finite difference scheme $\partial_\varphi r \simeq [r(x_0+\Delta x)-r(x_0-\Delta x)]/(2\Delta x) $.

\paragraph*{Measurement of the single-pixel sensitivity.} 
In order to measure the single-pixel sensitivity, we successively vary the phase shift $\varphi_j$ induced by each individual pixel $j$ for $100$ pixels covering an area of $6400$\,\textmu m$^2$ on the surface of the hidden SLM. This allows us to access the derivative of the outgoing field with respect to $\varphi_j$ using a centered difference scheme, and to calculate the associated Fisher information using Eq.~\eqref{eq_fisher_gaussian}. For each individual pixel, we perform an averaging of the derivative of the outgoing field over $10$ independent measurements. Once the Fisher information is calculated, we also subtract a residual noise floor that we estimate by taking different measurements of the same outgoing state. We perform the same analysis by illuminating the medium with the maximum information state and with different plane waves, and we normalize the values obtained for the maximum information state using the average value obtained with different plane waves. Note that a high single-pixel sensitivity cannot be achieved without a high intensity inside the pixel area. Thus, mapping the single-pixel sensitivity in the plane of the hidden SLM also provides us with an indirect way to approximate the intensity distribution in the plane of the hidden SLM.

\paragraph*{Monitoring of the global phase drift.}
Due to the significant acquisition time ($113$\,min in total), the global phase of the measured outgoing field slowly drifts in time because of the imperfect thermal stability of the experimental setup. During the acquisition, we continuously monitor this drift by regularly measuring a known outgoing field as a phase reference~\cite{pai_optical_2020}. We use different phase-reference fields depending on the incident power on the sample. 

When no density filters are placed in the optical path, the phase-reference field is generated by illuminating the medium with a given plane wave, with a slight angle so that no reflection from the back-focal plane of the objective can be observed. We then calculate how the global phase of this field changes over time by using a complex inner product of the phase-reference field measured at a given time with the phase-reference field measured at the beginning of the acquisition.

When any density filter is placed in the optical path, we first calculate a truncated reflection matrix $r'$, which does not include the few columns for which reflection from the back-focal plane of the objective can be observed. The phase-reference field is generated by illuminating the medium using the right-singular vector of $r'$ associated with its largest singular value. By doing so, we maximize the signal-to-noise ratio of phase-reference measurements. We then calculate how the global phase of this field changes over time by using a complex inner product of the phase-reference field measured at a given time with the field predicted from the knowledge of the reflection matrix.

Finally, we perform linear interpolations for estimating the global phase drift at any time during the acquisition, and we subsequently apply the appropriate phase correction to any measured data.


\begin{figure*}[!t]
		\begin{center}
			\includegraphics[width=9.5cm]{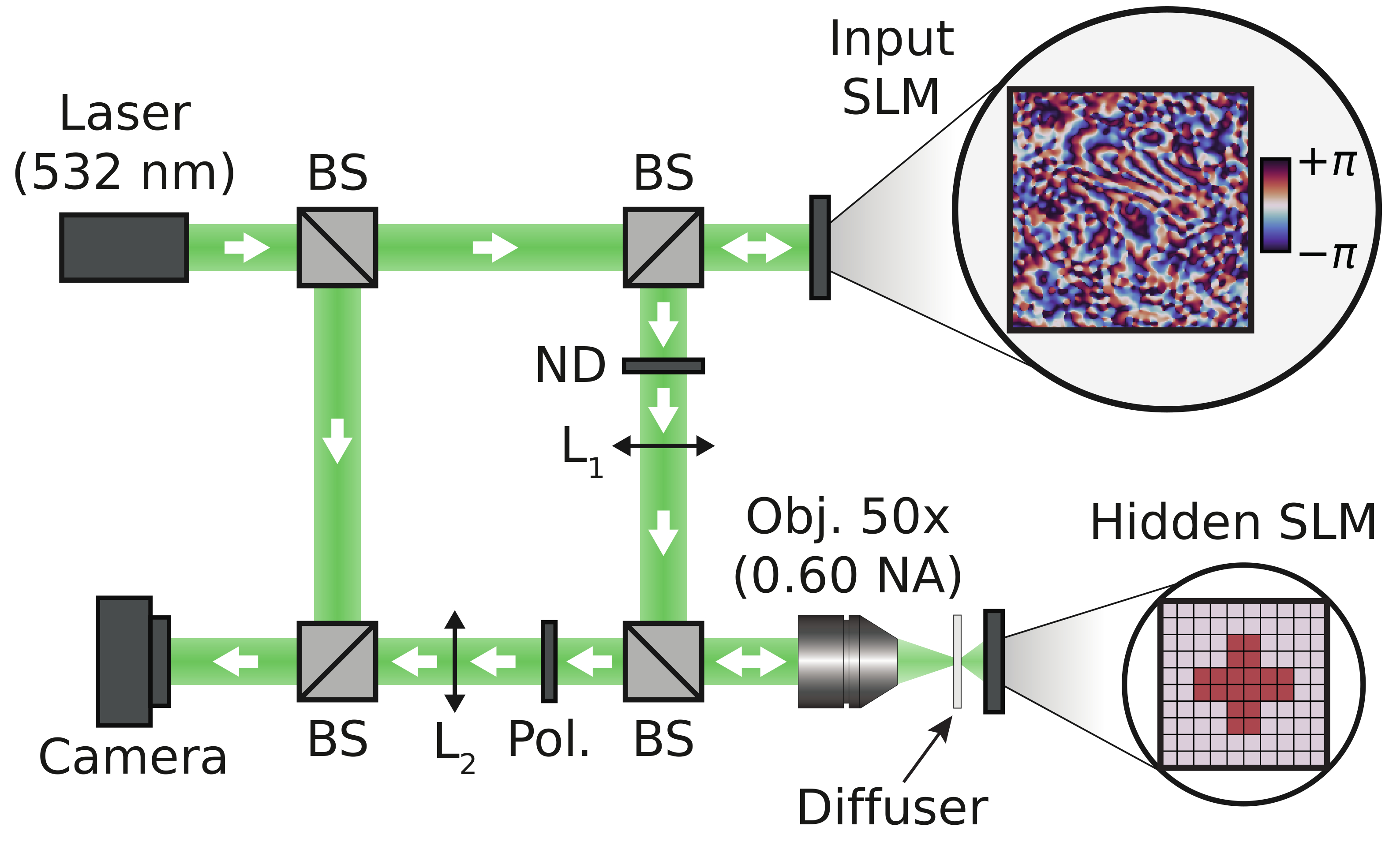}
		\end{center}
	\caption{\textbf{Extended Data Fig.~1. | Optical setup.} 
	The phases on the input SLM are modulated to reproduce the maximum information state, which reaches optimal sensitivity in its output with respect to any specified parameter characterizing the object displayed on the hidden SLM, such as phase variations or lateral displacements. Neutral density filters are removed for reflection matrix measurements. BS, beamsplitter; ND, neutral density filters; Obj, objective; NA, numerical aperture; Pol, linear polarizer; L$_1$ and L$_2$, lenses with focal length $200$\,mm.\vspace{5cm}}
\end{figure*}
\vspace{\fill}

\begin{figure*}[!t]
	\begin{center}
	\includegraphics[width=16.5cm]{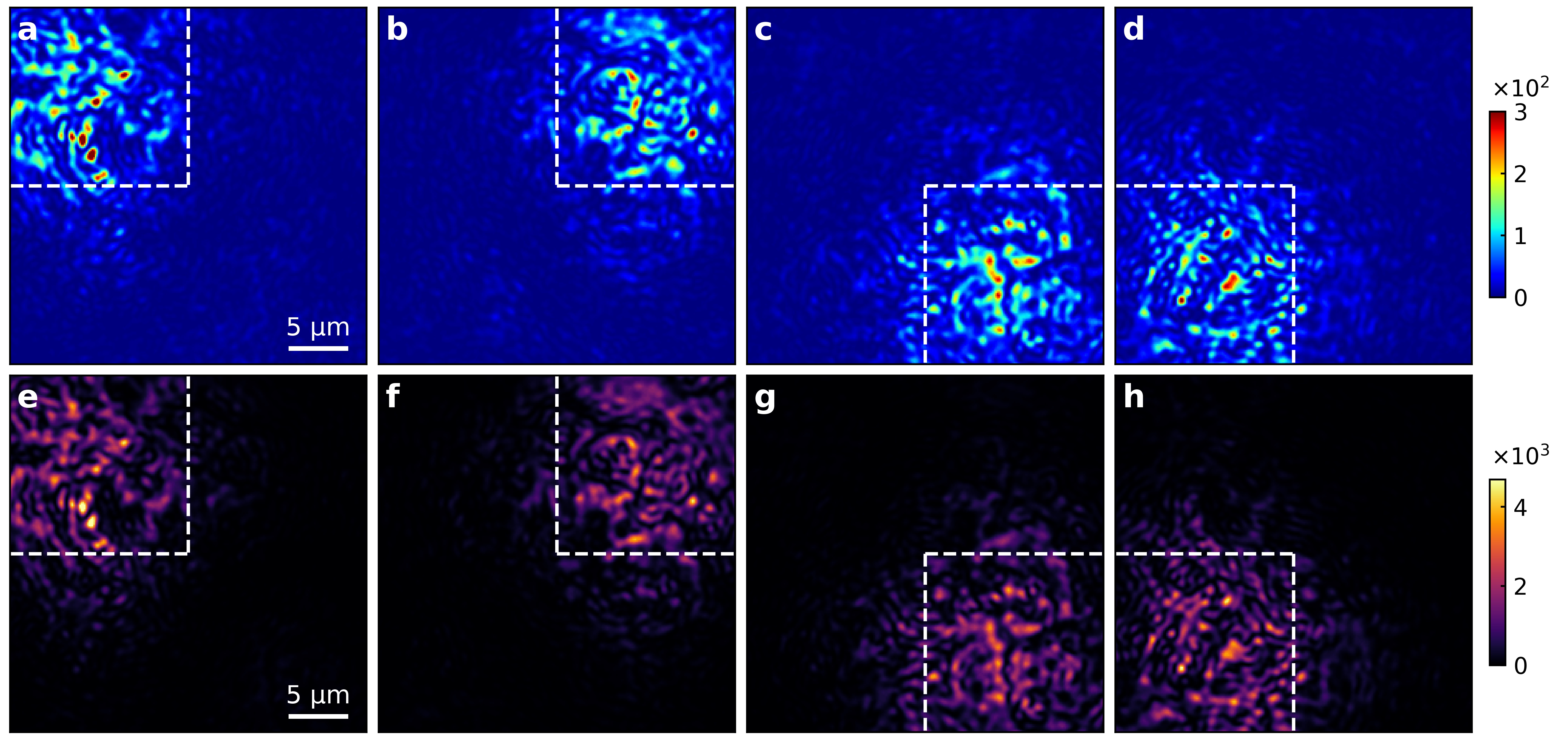}
	\end{center}
	\caption{\textbf{Extended Data Fig.~2. | Measured intensity and Fisher information distributions for maximum information states associated with different detection areas.}
	\textbf{a}--\textbf{d},~Measured spatial distributions of the intensity for optimal states, normalized by the average signal intensity under plane-wave illumination. The observable parameter is the phase shift induced by the cross-shaped target, and optimal states are defined here with respect to a reduced field of view of the camera that covers an area of $220$\,\textmu m$^2$, as delimited by white dashed lines. \textbf{e}--\textbf{h},~Analogous to \textbf{a}--\textbf{d} for the measured spatial distribution of the Fisher information per unit area. Remarkably, the maximum information state always delivers the Fisher information to the designated observer window.\vspace{5cm}}
\end{figure*}
\vspace{\fill}

\begin{figure*}[!t]
	\begin{center}
	\includegraphics[width=11cm]{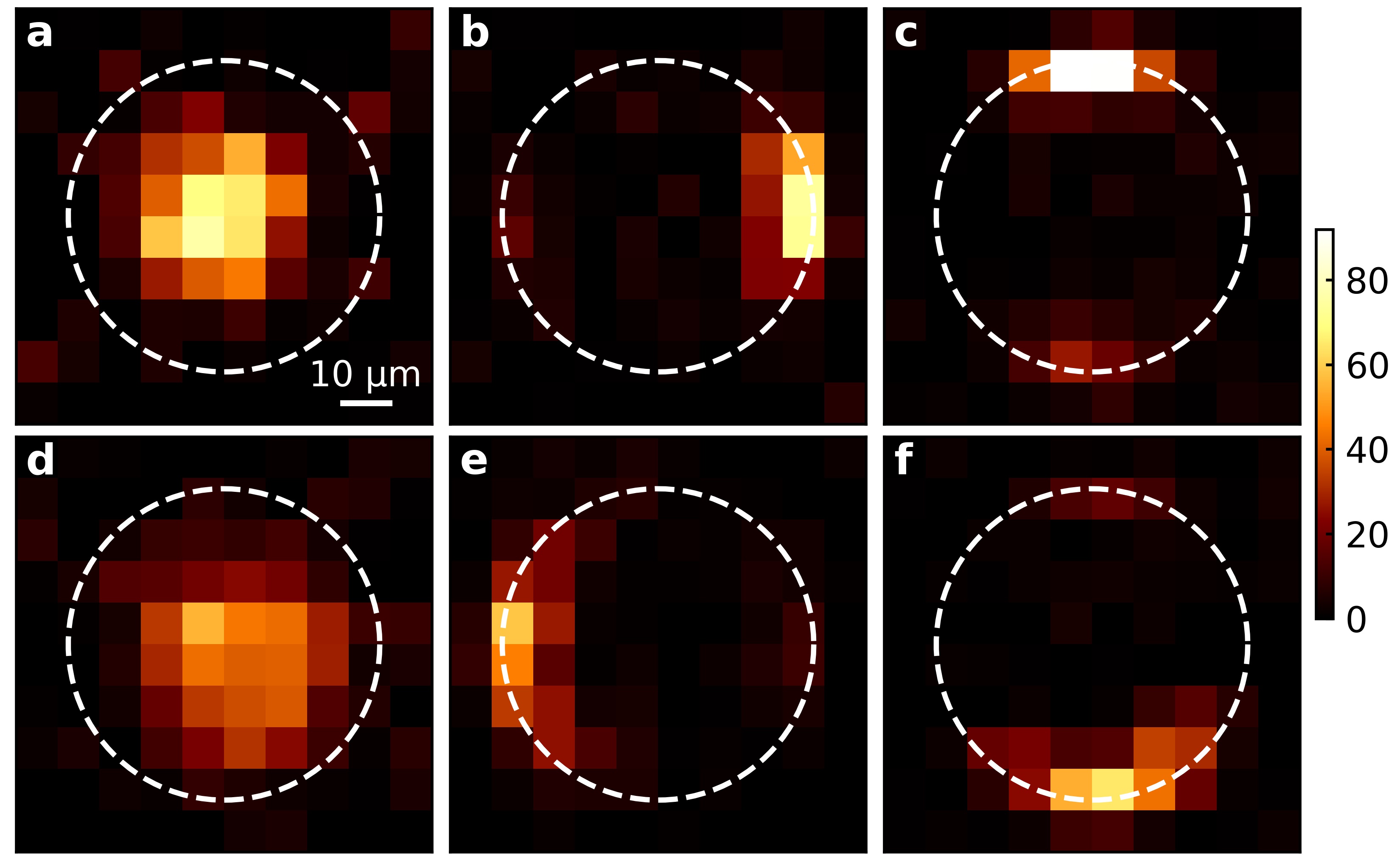}
	\end{center}
	\caption{\textbf{Extended Data Fig.~3. | Single-pixel sensitivity for maximum information states associated with different observable parameters.}
	\textbf{a}--\textbf{c},~Single-pixel sensitivity measured by shifting the phase of each pixel in the target area of the hidden SLM for the optimal state, normalized by the average single-pixel sensitivity under plane-wave illumination. The object displayed on the hidden SLM is a circular phase object, whose position is delimited by a white dashed circle. The field of view of the detection camera covers here an area of $880$\,\textmu m$^2$. The incident states used to illuminate the scattering medium are the maximum information states relative to a phase shift (\textbf{a}), a horizontal shift (\textbf{b}) and a vertical shift (\textbf{c}) of the object. \textbf{d}--\textbf{f},~Analogous to \textbf{a}--\textbf{c} when the field of view of the camera covers a reduced area of $144$\,\textmu m$^2$. In all cases, the maximum information state directs the incoming intensity to those parts on the hidden SLM that are most affected by the change in the observable parameter. Interestingly, when the target parameter is either a horizontal or a vertical shift of the object, the maximum information state typically focuses on a single edge rather than on both edges simultaneously, which is to be expected as the mirror symmetry in the system is broken by the diffuser.}
\end{figure*}
\vspace{\fill}

\begin{figure*}[!t]
		\begin{center}
			\includegraphics[width=10cm]{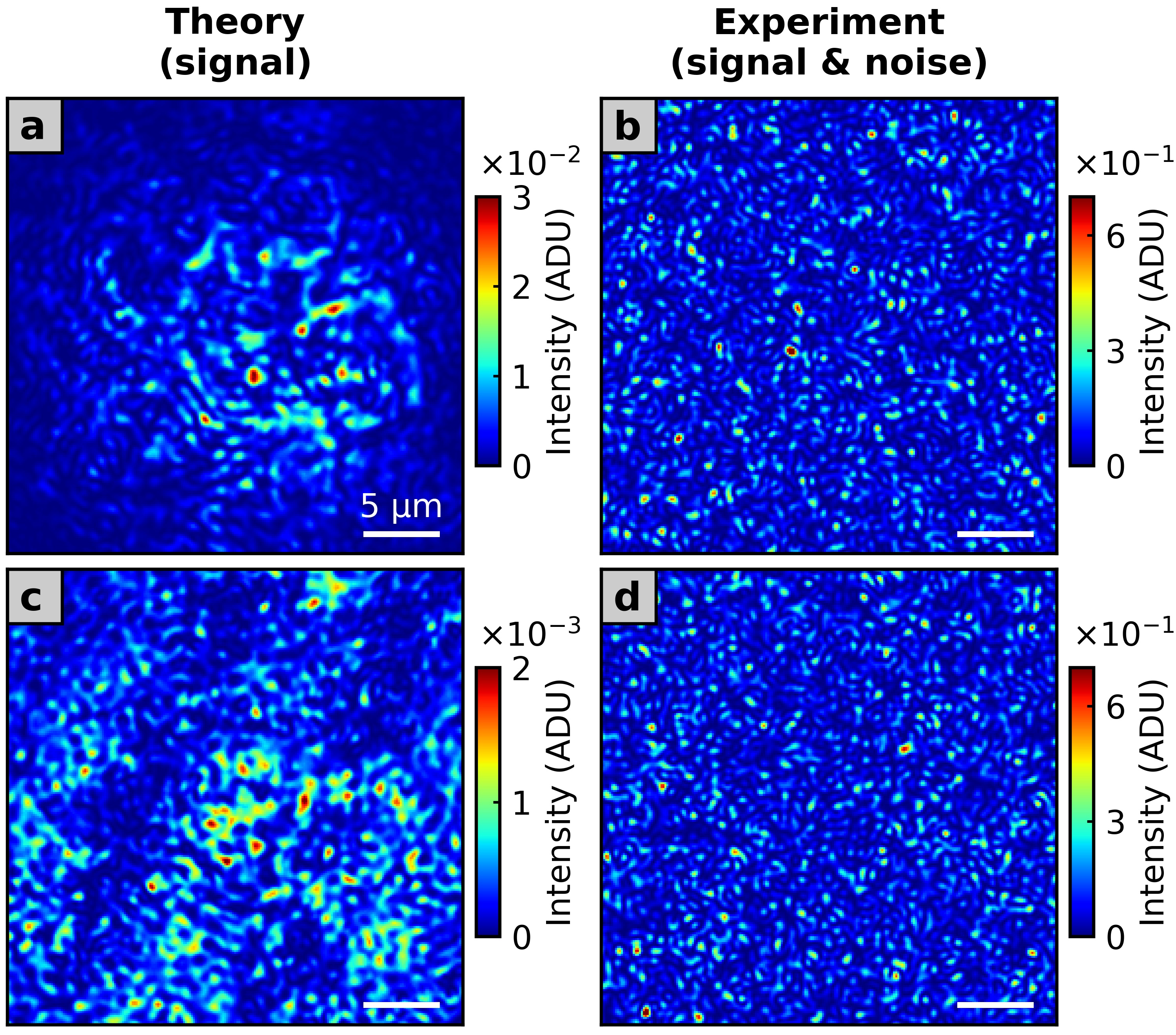}
		\end{center}
	\caption{\textbf{Extended Data Fig.~4. | Predicted and measured signal intensity distribution at low photon counts.}
	\textbf{a},~Predicted distribution of the signal intensity for the optimal incident state expressed in analog-to-digital units (ADU). This distribution is calculated from the measured reflection matrix, considering that the neutral density filter ND6 (fractional transmittance $8.3 \times 10^{-7}$) is placed in the optical path. \textbf{b},~Measured distribution of the signal intensity for the optimal incident state when ND6 is in the signal path. Such measurements are shot-noise limited, and the observed signal-to-noise per pixel is largely smaller than unity. Thus, the measured distribution of the signal intensity appears as a random noise, which has been low-pass filtered by the data analysis procedure used to digitally reconstruct complex fields from off-axis intensity measurements. Despite this low signal-to-noise per pixel, such data allow to correctly estimate the phase shift induced by the hidden target when using the minimum variance unbiased estimator. This can be achieved since only a single parameter (the phase shift induced by the target) needs to be estimated from a large number of independent sampling points. The Fisher information associated with each pixel of the detection camera effectively adds up, resulting in a total Fisher information that is sufficient to resolve the phase steps induced on the hidden SLM. \textbf{c}, \textbf{d},~Analogous to \textbf{a} and \textbf{b} for the best plane wave used to construct the reflection matrix.}
\end{figure*}
\vspace{\fill}

\begin{figure*}[!t]
		\begin{center}
			\includegraphics[width=0.5\linewidth]{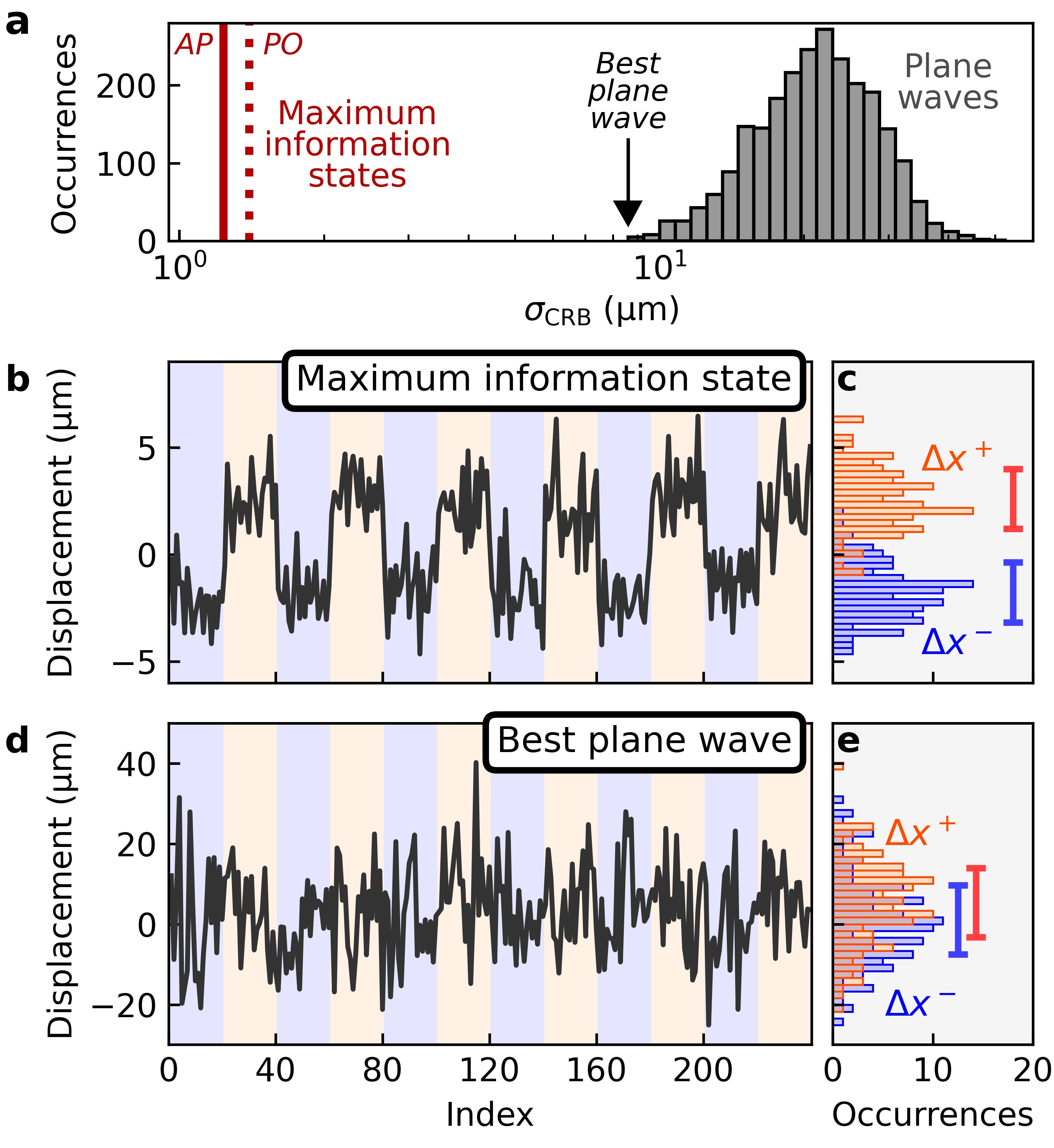}
		\end{center}
	\caption{\textbf{Extended Data Fig.~5. | Estimations of lateral displacements at low photon counts.} Analogous to Fig.~3 when the observable parameter is the horizontal position $x$ of the circular phase object shown in Fig.~2i.
	\textbf{a},~Histogram of precision limits for the $2437$ plane waves used to construct the reflection matrix and for the maximum information states (AP, amplitude and phase modulation; PO, phase-only modulation). \textbf{b},~Estimated lateral displacements of the circular phase object as a function of measurement index for measurements performed by illuminating the medium with the maximum information state. The calculated precision limit equals $1.4$\,\textmu m, and the observed standard error on the estimates is $1.5$\,\textmu m. \textbf{c},~Histograms of estimated angles for a positive lateral displacement ($\Delta x^+=+2.5$\,\textmu m) and a negative lateral displacement ($\Delta x^-=-2.5$\,\textmu m) applied by the hidden SLM. The length of error bars equals $2 \, \sigma_{\textsc{crb}}$. \textbf{d}, \textbf{e},~Analogous to \textbf{b} and \textbf{c} for measurements performed by illuminating the medium with the best plane wave.}
\end{figure*}
\vspace{\fill}


\onecolumngrid
\clearpage
\beginsupplement
\begin{center}
	\textbf{\large Maximum information states for coherent scattering measurements \\ \bigskip Supplementary information}
	
	\bigskip
	Dorian Bouchet,$^{1,2}$ Stefan Rotter,$^3$ and Allard P.\ Mosk$^1$\\ \vspace{0.15cm}
	\textit{\small $^\mathit{1}$Nanophotonics, Debye Institute for Nanomaterials Science and Center for Extreme Matter and Emergent Phenomena, Utrecht University, P.O. Box 80000, 3508 TA Utrecht, Netherlands \\ \smallskip
	\textit{\small $^\mathit{2}$Present address: Université Grenoble Alpes, CNRS, LIPhy, 38000 Grenoble, France}\\ \smallskip
	\textit{\small $^\mathit{3}$Institute for Theoretical Physics, Vienna University of Technology (TU Wien), 1040 Vienna, Austria}}
\end{center}
\vspace{0.3cm}	

\section*{Supplementary Section S1 -- Fisher information for the optimal detection scheme}

In this section, we calculate the Fisher information $\mathcal{J}(\theta)$ for the optimal homodyne detection scheme, which we show to be equal to the quantum Fisher information $\mathcal{I}(\theta)$ for coherent states and statistically-independent noise fluctuations. We then define the Fisher information operator from the scattering matrix of the system, and we provide the expression of the minimum variance unbiased estimator. We demonstrate that maximum information states can also be iteratively identified by performing time-reversal of a small perturbation using phase conjugation. Finally, we show that the Fisher information operator can be used in the context of multiple parameter estimations, in order to maximize the trace of the Fisher information matrix. 

\subsection*{S1.1 -- Fisher information}
Let us describe the measured data by a $N$-dimensional random variable $X$ and a joint probability density function $p(X;\theta)$ parameterized by an arbitrary parameter $\theta$. In general, the Fisher information is expressed by~\cite{kay_fundamentals_1993_1}
\begin{equation}
\mathcal{J}(\theta)= \Ee \left( [ \partial_\theta \ln p(X;\theta) ]^2 \right)\; ,
\label{def_information}
\end{equation}
where $\Ee $ denotes the expectation operator acting over noise fluctuations. In the shot-noise limit, $p(X;\theta)$ can be obtained using two equivalent strategies~\cite{mandel_optical_1995_1}:
\begin{itemize}
	\item Using a fully quantized picture, we can consider that $p(X;\theta)$ is determined by the quantum nature of the field. We define $|\alpha_k \ket$ as being the single-mode coherent (Glauber) state for the $k$-th outgoing spatial mode. Superposing this state with a reference coherent state $|r_k\ket$, the occupation of the state follows a Poisson distribution of expectation value $| \alpha_k+r_k|^2$, where $\alpha_k$ and $r_k$ are the eigenvalues of the annihilation operator $\hat{a}$ for the states $|\alpha_k \ket$ and $|r_k \ket$, respectively.  
	\item Using a semi-classical picture, we can consider that $p(X;\theta)$ is determined by the quantum nature of the photodetection process. We define $E_k^\out$ as being the complex value of the classical field for the $k$-th outgoing spatial mode. Superposing this field with a reference field $E_k^\rf$, the signal measured by the photodetector follows a Poisson distribution of expectation value $|E_k^\out+E_k^\rf|^2$.
\end{itemize}
Since the annihilation operator $\hat{a}$ is also the complex amplitude operator for coherent states~\cite{mandel_optical_1995_1}, we can write $E_k^\out=\bra \alpha_k | \hat{a} | \alpha_k \ket =\alpha_k$ and $E_k^\rf=\bra r_k | \hat{a} | r_k \ket =r_k$, evidencing the equivalence between the two approaches.

Adopting the semi-classical notation and considering that noise fluctuations are statistically independent for any two different outgoing modes, the joint probability density function $p(X;\theta)$ is thus expressed by
\begin{equation}
p(X;\theta) = \prod_{k=1}^N e^{-|E_k^\out+E_k^\rf|^2} \, \frac{|E_k^\out+E_k^\rf|^{2 X_k}}{X_k!} \; .
\label{pdf_classical}
\end{equation}
Injecting this expression in \eq{def_information}, we obtain
\begin{equation}
\mathcal{J}(\theta)= \sum_{k=1}^N \cfrac{\left[\partial_\theta \left(|E_k^\out+E_k^\rf|^2\right)\right]^2}{|E_k^\out+E_k^\rf|^2} \; .
\end{equation}
For a high-intensity reference field which does not depend on $\theta$ (i.e., for $|E_k^\rf|^2 \gg |E_k^\out|^2$ and $\partial_\theta E_k^\rf=0$), this expression simplifies to 
\begin{equation}
\mathcal{J}(\theta)= \sum_{k=1}^N \cfrac{\left( \partial_\theta \left[ E_k^\out \left(E_k^\rf\right)^* + E_k^\rf \left(E_k^\out\right)^* \right] \right)^2}{|E_k^\rf|^2} \; .
\label{eq_fisher_interf}
\end{equation} 
We now introduce $E_k^\out = Q_k^\out + i P_k^\out $ and $E_k^\rf=|E_k^\rf| e^{i \phi_k}$. \Eq{eq_fisher_interf} becomes
\begin{equation}
\mathcal{J}(\theta)= 4 \sum_{k=1}^N \left[ (\partial_\theta Q_k^\out) \cos \phi_k + (\partial_\theta P_k^\out) \sin \phi_k \right]^2 \; .
\label{eq_fisher_expanded}
\end{equation}
For any integer $m$, the phase angles of the reference field which maximize \eq{eq_fisher_expanded} are given by $\phi_k^\mathrm{max} = \arg (\partial_\theta E_k^\out)+ m \pi$, and those which minimize \eq{eq_fisher_expanded} are given by $\phi_k^\mathrm{min} = \arg (\partial_\theta E_k^\out)+ (m+1/2) \pi$. Choosing $\phi_k^\mathrm{min}$ as phase angles of the reference field yields $\mathcal{J}(\theta)= 0$. In contrast, choosing $\phi_k^\mathrm{max}$ as phase angles of the reference field yields
\begin{equation}
\mathcal{J}(\theta)= 4 \sum_{k=1}^N \lv \partial_\theta E^\out_k \rv^2 \; .
\label{fisher_optim_ref}
\end{equation}
This expression can be identified as the Fisher information for a random vector composed of $N$ complex random variables whose real and imaginary parts are independent normally distributed random variables with variance $\sigma^2=1/4$.

\subsection*{S1.2 -- Quantum Fisher information}

The Fisher information $\mathcal{J}(\theta)$ sets a lower bound on the variance of unbiased estimators of $\theta$ for a definite measurement scheme (i.e. the homodyne scheme in our case). A more general lower bound exists, which applies to any quantum measurement described by a positive-operator-valued measure (POVM). This lower bound is given by the reciprocal of the quantum Fisher information $\mathcal{I}(\theta)$, which is defined by~\cite{helstrom_quantum_1969_1}
\begin{equation}
\mathcal{I}(\theta)=\tr (\rho_\out L_\out^2) \; ,
\label{quantum_fisher}
\end{equation}
where $\rho_\out$ is a density operator describing the quantum state of the system and $L_\out$ is the symmetrized logarithmic derivative of $\rho_\out$ with respect to $\theta$ defined as follows:
\begin{equation}
\rho_\out L_\out + L_\out \rho_\out = 2 \, \partial_\theta \rho_\out \; .
\end{equation}
The Fisher information $\mathcal{J}(\theta)$ and the quantum Fisher information $\mathcal{I}(\theta)$ satisfy the inequality $\mathcal{I}(\theta)\geq \mathcal{J}(\theta)$, which is saturated when the POVM considered for the calculation of $\mathcal{J}(\theta)$ is optimal.

In our model, $\rho_\out$ describes an $N$-mode coherent state composed of simply-separable pure states, and thus \eq{quantum_fisher} simplifies to~\cite{braunstein_generalized_1996_1,demkowicz-dobrzanski_chapter_2015_1}
\begin{equation}
\mathcal{I}(\theta)= 4 \sum_{k=1}^N \left( \bra \partial_\theta \alpha_k | \partial_\theta \alpha_k \ket - |\bra \partial_\theta \alpha_k| \alpha_k \ket|^2 \right) \; ,
\label{quantum_fisher_pure}
\end{equation}
where $|\alpha_k \ket$ is the single-mode coherent state associated with the $k$-th mode and $| \partial_\theta \alpha_k \ket $ is its derivative with respect to $\theta$. We can represent $|\alpha_k \ket$ in the basis of Fock states $|n\ket$ labeled by the occupation number $n$, which reads~\cite{mandel_optical_1995_1}
\begin{equation}
|\alpha_k \ket = e^{- |\alpha_k|^2/2} \sum_{n=0}^{\infty} \frac{\alpha_k^n}{(n!)^{1/2}} |n\ket \; ,
\label{coherent_state}
\end{equation}
where $\alpha_k$ denotes the eigenvalue of the annihilation operator. Taking the derivative of this expression with respect to $\theta$ leads to
\begin{equation}
|\partial_\theta \alpha_k \ket = -\re \left( \alpha_k \partial_\theta \alpha_k^* \right) 
e^{- |\alpha_k|^2/2} \sum_{n=0}^{\infty} \frac{\alpha_k^n}{(n!)^{1/2}} |n\ket + \partial_\theta \alpha_k e^{- |\alpha_k|^2/2} \sum_{n=1}^{\infty} \frac{n\alpha_k^{n-1}}{(n!)^{1/2}} |n\ket  \; .
\label{der_coherent_state}
\end{equation}
We can now use \eqs{coherent_state}{der_coherent_state} to calculate the two terms $\bra \partial_\theta \alpha_k | \partial_\theta \alpha_k \ket$ and $|\bra \partial_\theta \alpha_k | \alpha_k \ket|^2$ that appear in \eq{quantum_fisher_pure}. Let us first calculate $\bra \partial_\theta \alpha_k | \partial_\theta \alpha_k \ket$ from \eq{der_coherent_state}: using the orthonormality of Fock states, we obtain
\begin{equation}
\begin{split}
\bra \partial_\theta \alpha_k | \partial_\theta \alpha_k \ket =&   \re  \left( \alpha_k \partial_\theta \alpha_k^* \right) ^2
e^{- |\alpha_k|^2} \sum_{n=0}^{\infty} \frac{|\alpha_k|^{2n}}{n!}  -  \re \left( \alpha_k \partial_\theta \alpha_k^* \right) \alpha_k \partial_\theta \alpha_k^* e^{- |\alpha_k|^2} \sum_{n=1}^{\infty} \frac{|\alpha_k|^{2(n-1)}}{(n-1)!}  \\
&  -  \re \left( \alpha_k \partial_\theta \alpha_k^* \right) \alpha_k^* \partial_\theta \alpha_k e^{- |\alpha_k|^2} \sum_{n=1}^{\infty} \frac{|\alpha_k|^{2(n-1)}}{(n-1)!} + |\partial_\theta \alpha_k|^2 e^{- |\alpha_k|^2} \sum_{n=1}^{\infty} \frac{n |\alpha_k|^{2(n-1)}}{(n-1)!} .
\end{split}
\label{term1_expanded}
\end{equation}
This expression can be simplified by using the following properties of exponential series:
\begin{equation}
\sum_{n=0}^\infty \frac{|\alpha_k|^{2n}}{n!} = e^{|\alpha_k|^2} \; ,
\label{expseries1}
\end{equation}
\begin{equation}
\sum_{n=0}^\infty \frac{(n+1)|\alpha_k|^{2n}}{n!} = (1+|\alpha_k|^2) e^{|\alpha_k|^2} \; .
\label{expseries2}
\end{equation}
Injecting \eqs{expseries1}{expseries2} into \eq{term1_expanded} leads to
\begin{equation}
\bra \partial_\theta \alpha_k | \partial_\theta \alpha_k \ket = | \partial_\theta \alpha_k |^2 + \im (\alpha_k \partial_\theta \alpha_k^*)^2 \; .
\label{first_term_coherent}
\end{equation}
Let us now calculate $\bra \partial_\theta \alpha_k | \alpha_k \ket$ from \eqs{coherent_state}{der_coherent_state}: using again the orthonormality of Fock states, we obtain
\begin{equation}
\bra \partial_\theta \alpha_k | \alpha_k \ket =  - \re \left( \alpha_k \partial_\theta \alpha_k^* \right)  
e^{- |\alpha_k|^2} \sum_{n=0}^{\infty} \frac{|\alpha_k|^{2n}}{n!} +  \alpha_k \partial_\theta \alpha_k^* e^{- |\alpha_k|^2} \sum_{n=1}^{\infty} \frac{|\alpha_k|^{2(n-1)}}{(n-1)!} \; .
\label{term2_expanded}
\end{equation}
Injecting \eq{expseries1} into \eq{term2_expanded} leads to
\begin{equation}
|\bra \partial_\theta \alpha_k | \alpha_k \ket|^2 = \im (\alpha_k \partial_\theta \alpha_k^*)^2  \; .
\label{second_term_coherent}
\end{equation}
Finally, injecting \eqs{first_term_coherent}{second_term_coherent} into \eq{quantum_fisher_pure} results in
\begin{equation}
\mathcal{I}(\theta) = 4 \sum_{k=1}^N | \partial_\theta \alpha_k |^2 \; .
\label{fisher_quantum_final}
\end{equation}
Recalling that $\alpha_k=E_k^\out$, the Fisher information $\mathcal{J}(\theta)$ given by \eq{fisher_optim_ref} equals the quantum Fisher information $\mathcal{I}(\theta)$ given by \eq{fisher_quantum_final}, thereby demonstrating that the homodyne detection scheme that we considered in Section S1.1 constitutes the optimal POVM for the estimation of $\theta$.

\subsection*{S1.3 -- Fisher information operator}
In the formalism of the $S$-matrix, the outgoing field state is expressed by $|E^\out\ket=S |E^\ins\ket$, where $S$ denotes the scattering matrix of the system. (Note that the incident state $|E^\ins\ket$ and the outgoing state $|E^\out\ket$ are defined here in the Hilbert space of all incident and outgoing spatial modes, whereas $|\alpha_k\ket$ introduced in Section S1.1 was defined in the Hilbert space of all quantum states for the $k$-th outgoing spatial mode.) We can thus write $E_k^\out$ as a projection of the outgoing state on the state associated with the $k$-th mode, which in bra-ket notation reads $E_k^\out=\bra k | S | E^\ins \ket$, leading to
\begin{equation}
\mathcal{J}(\theta)= 4 \sum_{k=1}^N \lv \bra k | \partial_\theta S | E^\ins \ket \rv^2 \; .
\end{equation}
This expression can be expanded into 
\begin{equation}
\mathcal{J}(\theta)= 4 \sum_{k=1}^N \bra E^\ins | (\partial_\theta S) ^\dagger |k \ket \bra k | \partial_\theta S | E^\ins \ket \; .
\end{equation}
Using the completeness relation $\sum_k \lv k \ket \bra k \rv =I_N$ where $I_N$ is the $N$-dimensional identity matrix, we finally obtain
\begin{equation}
\mathcal{J}(\theta)= 4 \ \bra E^\ins | (\partial_\theta S) ^\dagger \partial_\theta S | E^\ins \ket \; .
\label{eq_fisher_smatrix}
\end{equation}
In this expression, we can identify the operator $F_\theta = (\partial_\theta S) ^\dagger \partial_\theta S $, which we refer to as \textit{Fisher information operator}. We obtain the quadratic form $\mathcal{J}(\theta)= 4 \bra E^\ins | F_\theta | E^\ins \ket $, which is the expression of the Fisher information given in the manuscript. Finally, for a unitary scattering matrix ($S^\dagger=S^{-1}$), the operator $F_\theta$ is expressed by
\begin{equation}
F_\theta=(-i S^{-1} \partial_\theta S)^2 \; ,
\end{equation}
where we used the identity $\partial_\theta S^{-1} = - S^{-1} (\partial_\theta S) S^{-1}$. Introducing the generalized Wigner-Smith operator~\cite{ambichl_focusing_2017_1}, which is defined by $Q_\theta=-iS^{-1} \partial_\theta S$, we obtain the identity 
\begin{equation}
F_\theta= Q_\theta^2 \; ,  
\end{equation}
which implies that $F_\theta$ and $Q_\theta$ share the same eigenstates. Writing the eigenvalue equation $Q_\theta |\mathcal{E}^\ins_j\ket = \Theta_j |\mathcal{E}^\ins_j\ket$ where $\Theta_j$ is the $j$-th eigenvalue of $Q_\theta$ and $|\mathcal{E}^\ins_j\ket$ is the associated eigenstate, the outgoing field state satisfies $|\partial_\theta \mathcal{E}^\out_j \ket =i \Theta_j | \mathcal{E}^\out_j \ket $. For a scattering matrix evaluated at $\theta_0$ and a small parameter variation $\Delta \theta=\theta-\theta_0$, this results in $|\mathcal{E}^\out_j(\theta)\ket\simeq e^{i \Theta_j \Delta \theta} |\mathcal{E}^\out_j(\theta_0)\ket$. Thus, in the limit of a unitary scattering matrix, maximum information states are insensitive with respect to small variations in $\theta$ except for a global phase factor.

\subsection*{S1.4 -- Minimum variance unbiased estimator}

For small parameter variations around a given parameter value noted $\theta_0$, the measured data can be described by the following linear model:
\begin{equation}
X_k= I^{\tot}_k+ (\partial_\theta I^{\tot}_k ) (\theta-\theta_0) + W_k \; , 
\label{linear_model_intensity}
\end{equation}
where $X_k$ represents the intensity data measured by the camera at the $k$-th sampling point, where $I^{\tot}_k$ and $\partial_\theta I^{\tot}_k$ are evaluated at $\theta_0$, and where $W_k$ are $N$ independent and normally distributed random variables with mean zero and variance $I_k^{\tot}$. The normal distribution is indeed a good approximation of the Poisson distribution for large expectation values. In the case of linear models, general expressions exist for the minimum variance unbiased estimator, which depends on the noise statistics~\cite{kay_fundamentals_1993_1}. For the linear model expressed by \eq{linear_model_intensity}, the minimum variance unbiased estimator reads
\begin{equation}
\hat{\theta}(X) -\theta_0 = \frac{1}{\mathcal{J}(\theta_0)} \sum_{k=1}^N \frac{(\partial_\theta I_k^{\tot})(X_k-I_k^{\tot}) }{I_k^{\tot}} \; ,
\end{equation}
where $\hat{\theta}(X)$ denotes the estimator of $\theta$ (i.e. the function devised to estimate $\theta$ from the measured data $X$). Writing $I_k^{\tot}=|E_k^\out+E_k^\rf|^2$ and recalling that $|E_k^\rf|^2 \gg |E_k^\out|^2$ and $\partial_\theta E_k^\rf=0$, this expression simplifies to 
\begin{equation}
\hat{\theta}(X) -\theta_0 = \frac{1}{\mathcal{J}(\theta_0)} \sum_{k=1}^N \cfrac{ \left(\partial_\theta \left[ E_k^\out \left(E_k^\rf\right)^* + E_k^\rf \left(E_k^\out\right)^* \right]\right) (X_k-|E_k^\out+E_k^\rf|^2) }{|E_k^\rf|^2} \; .
\end{equation}
Introducing $E_k^\rf=|E_k^\rf| e^{i \phi_k}$ and using the phase angles $\phi_k^\mathrm{max} = \arg (\partial_\theta E_k^\out)$ which maximize the Fisher information leads to
\begin{equation}
\hat{\theta}(X) -\theta_0 = \frac{2}{\mathcal{J}(\theta_0)} \sum_{k=1}^N \cfrac{ |\partial_\theta E_k^\out|(X_k-|E_k^\out+E_k^\rf|^2 ) }{|E_k^\rf| } \; .
\end{equation}
In this expression, $\mathcal{J}(\theta_0)$ is obtained from \eq{fisher_optim_ref}, $E_k^\out$ is obtained from $E_k^\out=\bra k |S|E^\ins\ket$ and $\partial_\theta E_k^\out$ is obtained from $\partial_\theta E_k^\out=\bra k |\partial_\theta S|E^\ins\ket$, with a scattering matrix~$S$ evaluated at~$\theta_0$.

\subsection*{S1.5 -- Finding maximum information states using phase conjugation}

Time reversal using phase conjugation is a well-known technique that allows for instance to focus waves~\cite{lerosey_focusing_2007_1,xu_time-reversed_2011_1,judkewitz_speckle-scale_2013_1} or to identify open channels~\cite{bosch_frequency_2016_1} in multiple scattering media. In some implementations, light waves are focused into a scattering medium by performing time-reversal of a perturbation using phase conjugation~\cite{zhou_focusing_2014_1,ma_time-reversed_2014_1,ruan_focusing_2017_1}. Here, we show that an iterative procedure based on such techniques actually converges toward maximum information states. Light waves are then focused onto those specific areas of the object that are most affected by the perturbation, in such a way that the Fisher information available to the observer is maximized. In order to demonstrate the link between maximum information states and time-reversed adapted perturbation, we must restrict the analysis to a $N \times N$ scattering matrix that satisfies the reciprocity relation $S^\mathsf{T}=S$ (or equivalently $S^\dagger=S^*$). The procedure then relies on an iterative approach used to calculate the incident state $|E^\ins_{(n)}\ket $ at the $n$-th iteration from measurements based on an incident state $|E^\ins_{(n-1)}\ket $. The medium is illuminated with the incident state $|E^\ins_{(n-1)}\ket $ and, before a perturbation $\Delta \theta$ has occurred, the measured outgoing state is $|E^\out_{(n-1)}(\theta)\ket=S(\theta)| E^\ins_{(n-1)}\ket $. For the same incident state, the outgoing state measured after the perturbation has occurred is $|E^\out_{(n-1)}(\theta+\Delta \theta)\ket=S(\theta+\Delta \theta)| E^\ins_{(n-1)}\ket $. The new incident state $|E^\ins_{(n)}\ket $ is then obtained by phase-conjugating the difference between these two outgoing states. In the limit of a small perturbation $\Delta \theta$, this leads to
\begin{equation}
|E^\ins_{(n)} \ket = A_{(n)}^{-1/2} \;  \partial_\theta S ^* | E^\ins_{(n-1)} \ket ^* \; ,
\label{trap_increment}
\end{equation}
where $A_{(n)}^{-1/2}$ is a (real-valued) normalization coefficient. Using the relation $S^*=S^\dagger$, we can write $\partial_\theta S ^*=\partial_\theta S ^\dagger$. Then, starting from a random incident state $| E^\ins_{(0)} \ket$ and after $n$ successive iterations, we obtain
\begin{equation}
|E^\ins_{(n)} \ket = B_{(n)}^{-1/2} \; \left[(\partial_\theta S)^\dagger \partial_\theta S \right]^{n/2} | E^\ins_{(0)} \ket \qquad \text{if $n$ is even,}
\label{trap_even}
\end{equation}
\begin{equation}
|E^\ins_{(n)} \ket = B_{(n)}^{-1/2} \; \left[(\partial_\theta S)^\dagger \partial_\theta S \right]^{{n/2-1}} \partial_\theta S^\dagger | E^\ins_{(0)} \ket^* \qquad \text{if $n$ is odd.}
\label{trap_odd}
\end{equation}
Choosing the normalization condition $\bra E^\ins_{(n)} |E^\ins_{(n)} \ket =1$, the (real-valued) normalization coefficient $B_{(n)}$ is expressed by
\begin{equation}
B_{(n)} = \bra E^\ins_{(0)} | \left[(\partial_\theta S)^\dagger \partial_\theta S \right]^{n}  | E^\ins_{(0)} \ket \; .
\end{equation}
When the medium is illuminated with the incident state $|E^\ins_{(n)} \ket$, the Fisher information associated with the resulting outgoing state reads
\begin{equation}
\mathcal{J}_{(n)} (\theta) = 4 \bra E^\ins_{(n)} | (\partial_\theta S)^\dagger \partial_\theta S   | E^\ins_{(n)} \ket \; .
\label{trap_fisher_n}
\end{equation}
Using either \eq{trap_even} if $n$ is even or \eq{trap_odd} if $n$ is odd, \eq{trap_fisher_n} becomes 
\begin{equation}
\mathcal{J}_{(n)} (\theta)= 4  \; \frac{ \bra E^\ins_{(0)} | [(\partial_\theta S)^\dagger \partial_\theta S ]^{n+1}  | E^\ins_{(0)} \ket }{\bra E^\ins_{(0)} | \left[(\partial_\theta S)^\dagger \partial_\theta S \right]^{n}  | E^\ins_{(0)} \ket} \; .
\label{trap_fisher_0}
\end{equation}
Any incident state can be decomposed in the orthonormal basis formed by the eigenstates of $F_\theta$. In this basis, the initial state $|E^\ins_{(0)} \ket$ is expressed by
\begin{equation}
|E^\ins_{(0)} \ket = \sum_{j=1}^N \gamma_j |\mathcal{E}_j^\ins\ket \; ,
\label{trap_initial}
\end{equation}
where $|\mathcal{E}^\ins_j\ket$ is the $j$-th eigenstate of $F_\theta$ and $\gamma_j=\bra  \mathcal{E}_j^\ins  |E^\ins_{(0)} \ket $. From \eq{trap_initial} and using the eigenvalue equation $F_\theta |  \mathcal{E}^\ins_j \ket = \Lambda_j | \mathcal{E}^\ins_j\ket$, \eq{trap_fisher_0} becomes
\begin{equation}
\mathcal{J}_{(n)}(\theta)= 4\, \cfrac{\sum_{j=1}^N |\gamma_j|^2 \Lambda_j^{n+1}}{\sum_{j=1}^N |\gamma_j|^2 \Lambda_j^n} \; .
\label{trap_fisher}
\end{equation}
This expression can also be written in the following form:
\begin{equation}
\mathcal{J}_{(n)}(\theta)= 4 \Lambda_\mathrm{max} \times  \cfrac{\sum_{j=1}^N |\gamma_j|^2 (\Lambda_j/\Lambda_\mathrm{max})^{n+1}}{\sum_{j=1}^N |\gamma_j|^2 (\Lambda_j/\Lambda_\mathrm{max})^n} \; ,
\end{equation}
where $\Lambda_\mathrm{max}$ is the largest eigenvalue of $F_\theta$. Assuming that $|\gamma_j|^2\neq0$ for the associated eigenstate, we end up with
\begin{equation}
\lim_{n \to + \infty} \mathcal{J}_{(n)} = 4 \Lambda_\mathrm{max} \; .
\end{equation}
Thus, performing time-reversal of a perturbation using phase conjugation allows one to iteratively identify the maximum information state relative to the perturbation, provided that the scattering matrix describing the medium is a square matrix satisfying the reciprocity condition $S^\mathsf{T}=S$, and that the overlap between the initial state and the maximum information state is not equal to zero. 

It clearly appears that the eigenstates of $F_\theta$ constitute a relevant basis to analyze this iterative procedure. Indeed, the number of iterations needed for the procedure to converge depends on both the values of $\gamma_j$ and the eigenvalue spectrum of $F_\theta$, according to \eq{trap_fisher}. In order to illustrate this feature, we consider the operator $f_\varphi=(\partial_\varphi r)^\dagger \partial_\varphi r $ measured by shifting the phase $\varphi$ induced by the cross-shaped object. We numerically generate $10\,000$ random initial states and we use \eq{trap_fisher} to calculate the ratio $\mathcal{J}_{(n)}(\varphi)/\mathcal{J}_\mathrm{max}(\varphi)$, where $\mathcal{J}_\mathrm{max}(\varphi)$ is the Fisher information associated with the maximum information state. After the first iteration, the median value of the ratio $\mathcal{J}_{(n)}(\varphi)/\mathcal{J}_\mathrm{max}(\varphi)$ is equal to $0.42$ (Fig.~S1), and we observe that $4$ iterations are generally required to identify a state that reaches $90\%$ of the optimal Fisher information. Moreover, some initial states show a small overlap with the maximum information state; for such states, the ratio $\mathcal{J}_{(n)}(\varphi)/\mathcal{J}_\mathrm{max}(\varphi)$ after $10$ iterations is close to $0.6$, corresponding to the second largest eigenvalue of $f_\varphi$.

\begin{figure}[ht]
	\begin{center}
		\includegraphics[width=8.5cm]{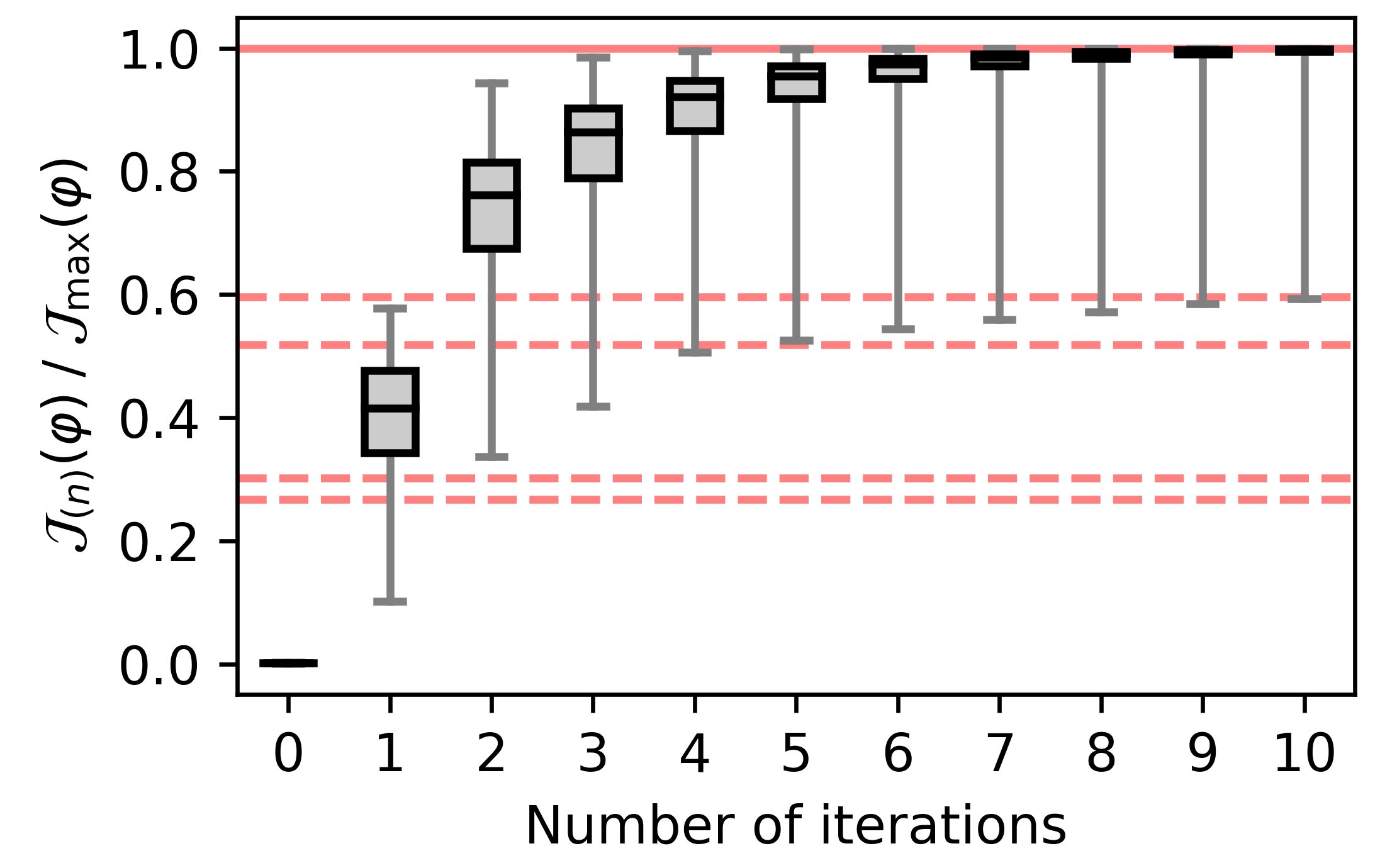}
	\end{center}
	\caption{\textbf{Supplementary Figure S1 | Fisher information maximization using time-reversed adapted perturbation.}
	Fisher information associated with light states identified using time-reversed adapted perturbation for $10\,000$ random initial states, normalized by the Fisher information of the maximum information state. A horizontal line goes through each box at the median value, edges of the boxes represent lower and upper quartiles, and whiskers represent minimum and maximum values. The box and whiskers for $0$ iterations appear as a single horizontal line; indeed, random initial states generate a Fisher information that is much smaller than the maximum information state.  The solid red line represents the Fisher information for the largest eigenvalue of $f_\varphi$, and the dashed red lines represent the Fisher information for the second, third, fourth and fifth largest eigenvalues.}
\end{figure}

\subsection*{S1.6 -- Multiple parameter estimations}

When several parameters $\theta=(\theta_1,\dots,\theta_p)^{\mathsf{T}}$ need to be simultaneously estimated, the Cramér-Rao inequality reads~\cite{kay_fundamentals_1993_1}
\begin{equation}
\operatorname{Var} \left( \hat{\theta}_i \right) \geq \left[ \mathcal{J}^{-1}(\theta)\right]_{ii} \; ,
\end{equation}
where $\hat{\theta}_i$ is any unbiased estimator of $\theta_i$, $\operatorname{Var}$ denotes the variance operator and $\mathcal{J}(\theta)$ is now a $p \times p$ Fisher information matrix defined as follows:
\begin{equation}
\left[\mathcal{J}(\theta)\right]_{ij} = \Ee \, \left( [ \partial_{\theta_i} \ln p(X;\theta) ] [ \partial_{\theta_j} \ln p(X;\theta)] \right) \,.
\end{equation}
Diagonal elements of the Fisher information matrix can then be interpreted as the amount of information relevant to the estimation of each parameter taken individually, while off-diagonal elements describe the influence of possible correlations between estimated values of different parameters. 

For scattering measurements performed with the optimal homodyne detection scheme, the trace of the Fisher information matrix is expressed by
\begin{equation}
\operatorname{Tr} \left[ \mathcal{J}(\theta) \right] = 4  \bra E^\ins | \sum_{i=1}^p F_{\theta_i} | E^\ins  \ket \; ,
\end{equation}
where $F_{\theta_i}= (\partial_{\theta_i} S) ^\dagger \partial_{\theta_i} S $ is the Fisher information operator associated with the $i$-th parameter. Thus, the incident state that maximizes the trace of the Fisher information matrix is found by calculating the eigenstates of the operator $\sum_i F_{\theta_i}$ and by choosing the one associated with the largest eigenvalue. Whenever the Fisher information matrix is diagonal, we can write $[ \mathcal{J}^{-1}(\theta)]_{ii} = [ \mathcal{J}(\theta)]_{ii}^{-1} $. In this case, maximizing the trace of the Fisher information matrix improves the measurement precision for all parameters simultaneously. Furthermore, any weighted sum of diagonal elements of the Fisher information matrix can be maximized using the same approach, a feature which could be useful to improve the precision for specific parameters of interest. In contrast, if the Fisher information matrix is not diagonal, we can only write $[ \mathcal{J}^{-1}(\theta)]_{ii} \geq [ \mathcal{J}(\theta)]_{ii}^{-1} $. It is then still needed to maximize the value of diagonal elements but it is also important to minimize the value of off-diagonal elements, thus requiring a different optimization strategy. If an iterative algorithm is used for this purpose, maximizing the trace of the Fisher information matrix can yield a good initial guess for the algorithm.

\section*{Supplementary Section S2 -- Fisher information in the experiments}

In this section, we give the expression of the Fisher information for our optical setup and we provide the expression of the minimum variance unbiased estimator used in the experiments.

\subsection*{S2.1 -- Expression of the Fisher information in the experiments}

In our proof-of-principle experiments, the reference beam is not optimally shaped but consists of a tilted plane wave. In this case, the Fisher information is obtained by averaging \eq{eq_fisher_expanded} over the phase angle of the reference beam:
\begin{equation}
\mathcal{J}(\theta)= \frac{2}{\pi} \sum_{k=1}^N  \int_0^{2\pi} \de \phi \left[ (\partial_\theta Q_k^\out) \cos \phi + (\partial_\theta P_k^\out) \sin \phi\right]^2 \; .
\end{equation}
We obtain the following simplified expression:
\begin{equation}
\mathcal{J}(\theta)= 2 \sum_{k=1}^N \lv \partial_\theta E^\out_k \rv^2 \; .
\label{fisher_mean_ref}
\end{equation}
Note that this expression is identical to the expression of the optimal Fisher information given in \eq{fisher_optim_ref}, except for a factor of two. Hence, the incoming state which maximizes \eq{fisher_mean_ref} also maximizes \eq{fisher_optim_ref}.

In the experiments, we must also take into account two additional considerations. First, the value of the $N=2465$ sampling points are actually estimated from $N_\mathrm{cam}=53\,044$ statistically independent camera pixels (oversampling is required in off-axis holography). The oversampling ratio $\beta=N_\mathrm{cam}/N$ must therefore be included as a prefactor in \eq{fisher_mean_ref}. Moreover, while we measured the reflection matrix without any neutral density filter, we consistently use the strongest neutral density filter ND6 (fractional transmittance $\mathcal{T}=0.83 \times 10^{-6}$) to compute all values of the Fisher information. Therefore, the expression of the Fisher information relevant to our experiments is 
\begin{equation}
\mathcal{J}(\theta)= \frac{\mathcal{T}}{\sigma^2} \bra E^\ins | (\partial_\theta r) ^\dagger \partial_\theta r | E^\ins \ket \; ,
\label{Fisher_exp}
\end{equation}
where we noted $\sigma^2=1/(2\beta)$ in order to highlight that this expression can be identified as the Fisher information for a random vector composed of $N$ complex random variables whose real and imaginary parts are independent normally distributed random variables with variance $\sigma^2$. The precision limit $\sigma_{\textsc{crb}}$, which bounds the standard deviation of any estimator of $\theta$ in our experiments, is then expressed from \eq{Fisher_exp} by $\sigma_{\textsc{crb}}=\mathcal{J}^{-1/2}$. 

We can verify that the measured field quadratures follow a Gaussian distribution when the neutral density filter ND6 is in the signal path. To this end, we illuminate the medium with the optimal incident state relative to the estimation of $\varphi$ and we perform $100$ successive measurements of the outgoing field. We then compute the p-value for each sampling point according to the Shapiro-Wilk test. The uniform distribution of p-value (Fig.~S2a) confirms that the measured field quadratures follow a Gaussian distribution. We also construct the histogram of the sample variance (Fig.~S2b), with a mean value of $0.0243$\,ADU. This value is in good agreement with the theoretical value given by $1/(2\beta)=0.0232 \; \mathrm{ADU}$. We consistently choose the mean value experimentally determined ($\sigma^2=0.0243~$ADU) to calculate the Fisher information in \eq{Fisher_exp}. 

\begin{figure}[ht]
	\begin{center}
		\includegraphics[width=10.5cm]{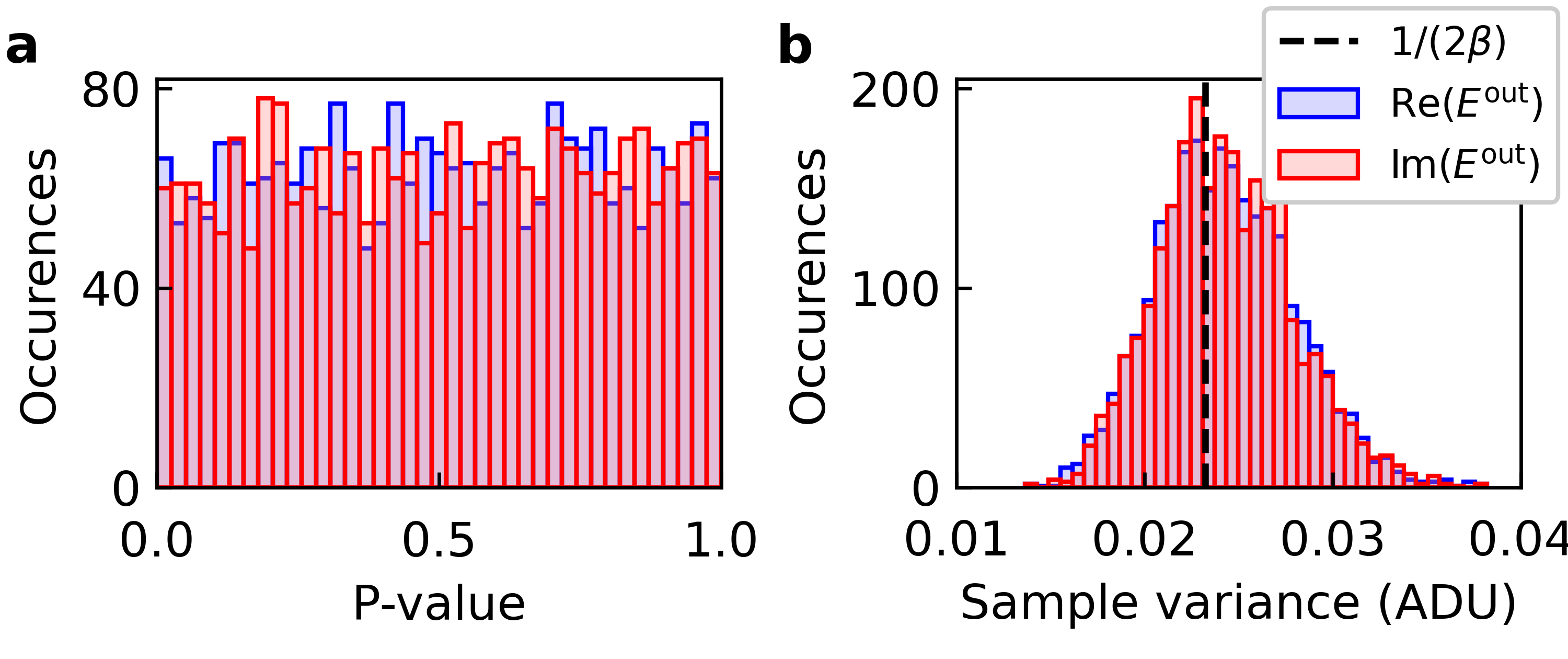}
	\end{center}
	\caption{\textbf{Supplementary Figure S2 | Statistics of the field quadratures.}
	\textbf{a}, P-value distribution for the $2465$ sampling points. The p-value for each point is calculated using Shapiro-Wilk test statistics from $100$ different realization of the random noise. \textbf{b},~Distribution of the sample variance for the $2465$ sampling points.}
\end{figure}

\subsection*{S2.2 -- Minimum variance unbiased estimator in the experiments}

For small variations of $\theta$ around $\theta_0$, measured data can be described by the following linear model:
\begin{equation}
Z_k= E^{\out}_k+ (\partial_\theta E^{\out}_k ) (\theta-\theta_0) + W_k \; , 
\end{equation}
where $Z_k$ represents here the complex field retrieved from the measured data using off-axis holography and evaluated at the $k$-th sampling point, where $E^{\out}_k$ and $\partial_\theta E^{\out}_k$ are evaluated at $\theta_0$, and where $W_k$ are $N$ independent complex random variables whose real and imaginary parts are independent normally distributed random variables with mean zero and variance $\sigma^2$. For this linear model, the minimum variance unbiased estimator reads~\cite{kay_fundamentals_1993_1}
\begin{equation}
\hat{\theta}(Z) -\theta_0 = \frac{\re \left[ \bra \partial_\theta E^\out \lv Z \ket -\bra \partial_\theta E^\out | E^\out \ket \right]}{\bra \partial_\theta E^\out | \partial_\theta E^\out \ket } \; .
\label{eq_estimator}
\end{equation}
All estimations presented in the manuscript are performed from data measured with the neutral density filter ND6 (characterized by a fractional transmittance $\mathcal{T}$) placed in the signal optical path. Moreover, due to measurement noise during reflection matrix measurements, the intensity and the Fisher information predicted from the knowledge of the reflection matrix differ from direct measurements by a factor $\eta_\textsc{i}$ and $\eta_\textsc{f}$, respectively (see Supplementary Information~S3). The estimator of $\theta$ associated with our experimental setup is thus obtained from \eq{eq_estimator} by taking $|E^\out\ket=(\eta_\textsc{i} \mathcal{T})^{1/2} \, r |\tilde{E}^\ins\ket$ and $ |\partial_\theta E^\out \ket= (\eta_\textsc{f} \mathcal{T})^{1/2} \, \partial_\theta r |\tilde{E}^\ins\ket$. We obtain the following expression:
\begin{equation}
\hat{\theta}(Z) - \theta_0 = \frac{\re \left[ (\mathcal{T} \eta_\textsc{f})^{-1/2} \bra \tilde{E}^\ins \rv (\partial_\theta r) ^\dagger \lv Z \ket - (\eta_\textsc{i}/ \eta_\textsc{f})^{1/2} \bra \tilde{E}^\ins \rv (\partial_\theta r) ^\dagger r \lv \tilde{E}^\ins \ket\right]}{ \bra \tilde{E}^\ins \rv \left(\partial_\theta r\right) ^\dagger \partial_\theta r \lv \tilde{E}^\ins \ket } \; .
\label{eq_estimator2}
\end{equation}
Evaluating this estimator using measured data $Z$ directly yields the estimates shown in the manuscript. Importantly, all experimental parameters involved in this expression (i.e. $\mathcal{T}$, $\eta_\textsc{i}$ and $\eta_\textsc{f}$) are characterized using independent measurements. Note that, whereas estimates shown in the manuscript are almost unbiased, small biases often appear when applying \eq{eq_estimator2} to experimental data. Such biases, which are usually smaller than $2\sigma_{\textsc{crb}}$, can be explained by the influence of measurement noise in the reflection matrices and by an imperfect correction of the global phase variations induced by thermal instabilities of the setup.

\section*{Supplementary Section S3 -- Characterization of the outgoing field and its derivative}

In this section, we show that reflection matrix measurements allow to faithfully predict the outgoing field and its derivatives with respect to the phase and to the position of hidden objects, and we characterize the correlation between the outgoing field distribution and its derivative.

\subsection*{S3.1 -- Predicting the outgoing field}

In general, illuminating the scattering medium with an arbitrary state $|E^\ins\ket$ requires amplitude and phase modulation of the incident field state. However, in the experiments, only the phase of the field can be modulated by the input SLM. We must therefore determine the expression of the phase-only modulated state $|\tilde{E}^\ins\ket$ that is experimentally used to illuminate the medium instead of $|E^\ins\ket$. Let us define the SLM pattern $|E_\textsc{slm}\ket=M |E^\ins\ket$, where $M$ is a transformation matrix mapping incident states (expressed in a basis of plane waves) to SLM patterns (expressed in a basis of SLM pixels). The phase-only modulated state can be numerically approximated by
$|\tilde{E}^\ins\ket = \mathrm{argmin} ( \Vert |\tilde{E}_\textsc{SLM}\ket - M |E^\ins\ket \Vert, |E^\ins\ket )$, where $|\tilde{E}_{\textsc{slm}}\ket$ is the SLM pattern that has the same phase as $|E_\textsc{slm}\ket$ but with uniform amplitude. Thus, predicted outgoing states are given by $|E^\out_{\mathrm{ap}}\ket = r |E^\ins\ket$ for an amplitude and phase modulation, and by $|E^\out_{\mathrm{po}}\ket = r |\tilde{E}^\ins\ket$ for a phase-only modulation.

\begin{figure}[b!]
	\begin{center}
		\includegraphics[width=14.6cm]{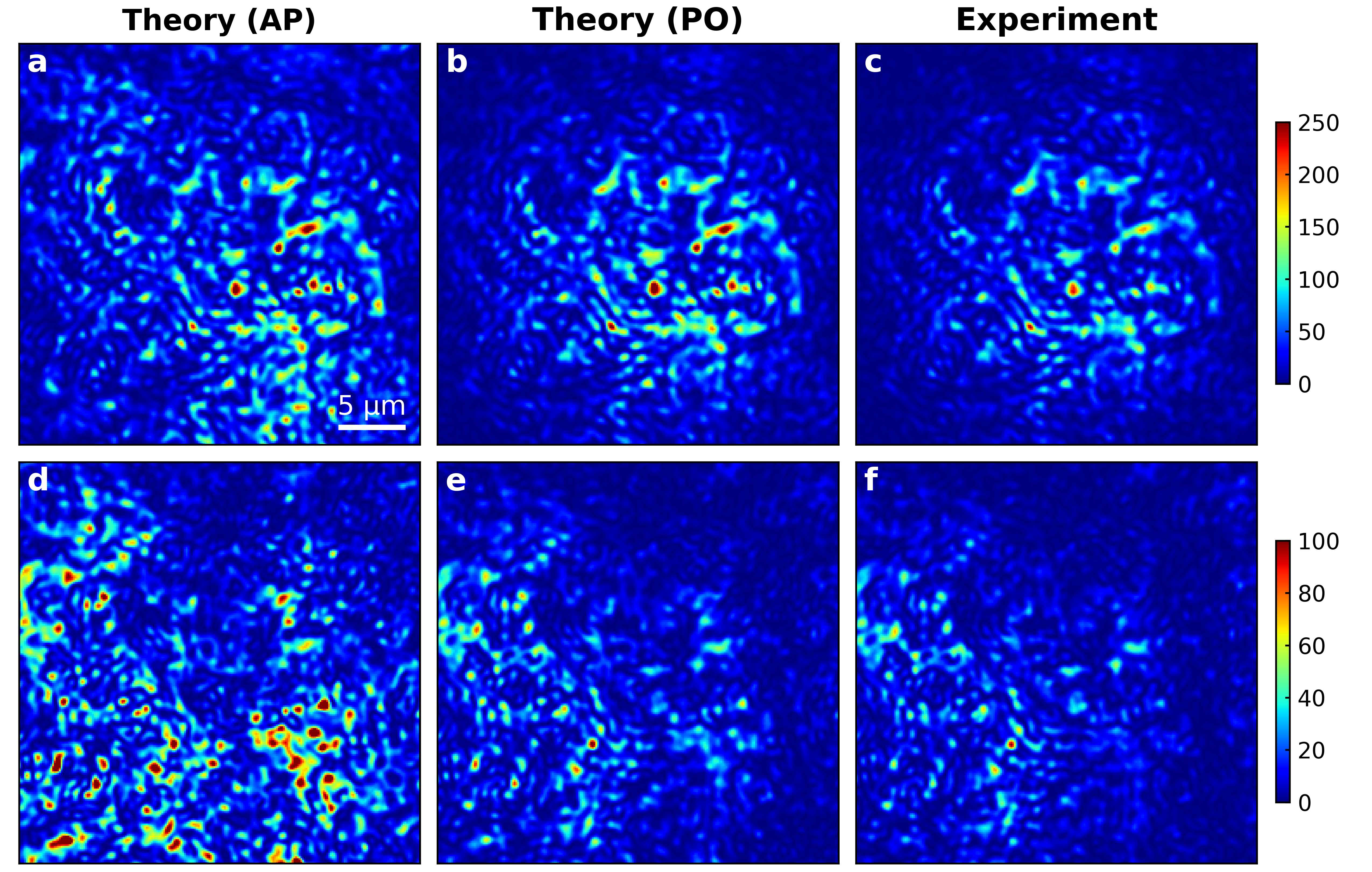}
	\end{center}
	\caption{\textbf{Supplementary Figure S3 | Predicted and measured intensity distributions for maximum information states.}
	\textbf{a}, \textbf{b},~Predicted spatial distribution of the normalized intensity for amplitude and phase (AP) modulation and phase-only (PO) modulation, when the observable parameter is the phase shift of the cross-shaped object. \textbf{c},~Measured spatial distribution of the normalized intensity. \textbf{d}--\textbf{f} Analogous to \textbf{a}--\textbf{c} when the observable parameter is the lateral displacement of the circular object.}
\end{figure}

We can test this procedure by measuring the outgoing field when the medium is illuminated using optimal incident states. This characterization is experimentally performed by averaging the outgoing field over $10$ measurements, with the neutral density filter ND2 placed in the signal path (measured fractional transmittance $\mathcal{T}'=0.76 \times 10^{-2}$) to avoid saturation of the camera. The complex correlation coefficients between measured and predicted fields are $0.96-0.21 \, i$ (for the maximum information state relative to the estimation of the phase shift of the cross-shaped object) and $0.95+0.06 \, i$ (for the maximum information state relative to the estimation of the lateral shift of the circular object). These values show that we can correctly predict the outgoing states from the measured reflection matrix. Comparing the spatial distributions of the measured intensity to the predicted ones, we can see that these distributions are indeed very similar (Fig.~S3). However, due to the presence of noise in reflection matrix measurements, we observe that the total measured intensity is lower by a factor of the order of $\eta_\textsc{i}\simeq0.80$ when compared to the predicted intensity.

\subsection*{S3.2 -- Predicting the derivative of the outgoing field}

In order to faithfully assess the Fisher information in the experiments, we must ensure that $\partial_\theta r$ is correctly estimated using the finite-difference scheme $\partial_\theta r \simeq [r(\theta_0+\Delta \theta) - r(\theta_0-\Delta \theta)]/( 2 \Delta \theta)$. To this end, the change in the measured outgoing states generated by parameter variations of $\pm \Delta \theta$ must be larger than the level of noise in the measurements. In Fig.~S4a, we consider the case in which the observable of interest is the phase shift induced by the cross-shape object, and we show the distribution of the estimated Fisher information for each incident plane wave used to generate the reflection matrix (blue histogram). For comparison purpose, we show the distribution of noise estimates (red histogram), obtained for each plane wave by taking two identical measurements of the outgoing state for $\varphi=\varphi_0$. We clearly observe that the signal is significantly larger than the noise, with a signal mean value of $0.41$\,rad$^{-2}$ and a noise mean value of $0.13$\,rad$^{-2}$. Similarly, in Fig.~S4b, we consider the case in which the observable of interest is the lateral displacement of the circular object, and we show the distribution of the estimated Fisher information for each incident plane wave used to generate the reflection matrix along with the distribution of noise estimates. In this case, the signal is also larger than the noise, with a signal mean value of $0.0027$\,\textmu m$^{-2}$ and a noise mean value of $0.0014$\,\textmu m$^{-2}$.

\begin{figure}[ht]
	\begin{center}
		\includegraphics[width=14.6cm]{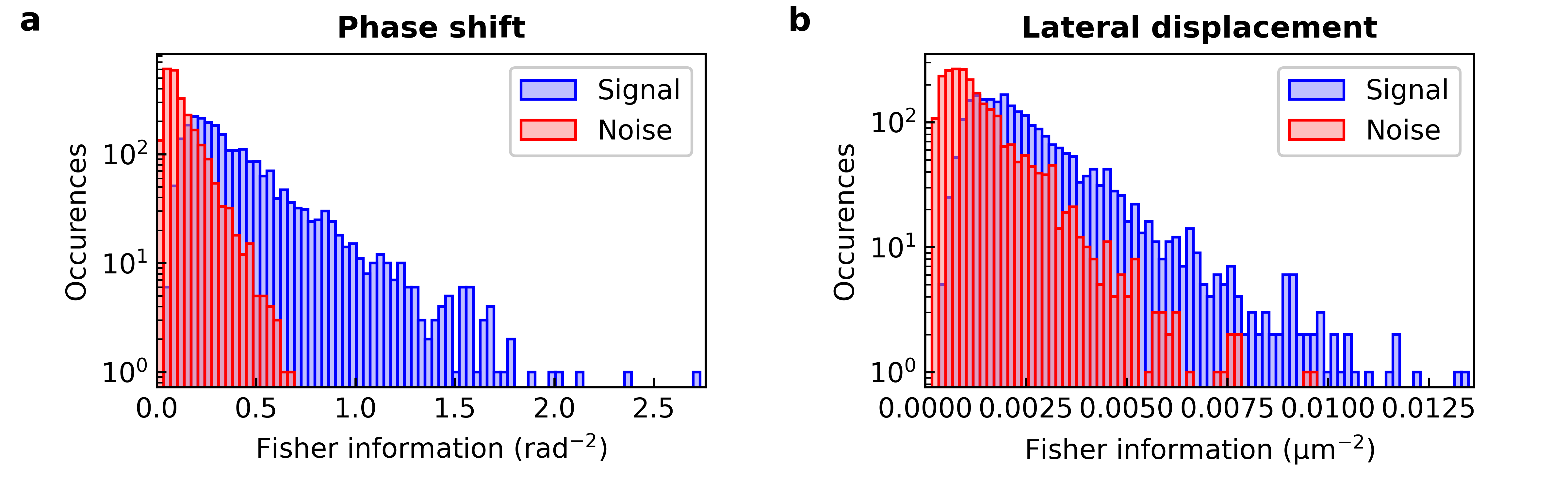}
	\end{center}
	\caption{\textbf{Supplementary Figure S4 | Histograms of Fisher information and measurement noise.}
	\textbf{a},~Histogram of Fisher information for the $2437$ plane waves used to construct the reflection matrix, along with the histogram of associated measurement noise, when the observable parameter is the phase shift of the cross-shaped object. \textbf{b},~Analogous to \textbf{a} when the observable parameter is the lateral displacement of the circular object.}
\end{figure}

We can verify that this signal-to-noise ratio is sufficient to faithfully estimate $\partial_\varphi r$ and $\partial_x r$ by measuring the derivative of the outgoing field when the optimal incident state is used to illuminate the medium, and by comparing it to the derivative of the field predicted using the derivative of the reflection matrix. This characterization is experimentally performed by averaging the derivative of the outgoing field over $10$ measurements, with ND2 in the signal path to avoid saturation of the camera. The complex correlation coefficients between measured and predicted derivative of the field are $0.97-0.22 \, i$ (for the maximum information state relative to the estimation of the phase shift of the cross-shaped object) and $0.99+0.00 \, i$ (for the maximum information state relative to the estimation of the lateral shift of the circular object). These values show that we can correctly predict the derivative of the outgoing field from reflection matrix measurements. Comparing the spatial distributions of the measured Fisher information per unit area to the predicted ones, we can see that these distributions are very similar (Fig.~S5). However, the total measured Fisher information is lower by a factor of the order of $\eta_\textsc{f}\simeq 0.75$ when compared to the predicted Fisher information. This difference is largely explained by the observed difference in the predicted and measured outgoing intensity (for which a factor $\eta_\textsc{i}\simeq0.80$ was measured).

\begin{figure}[ht]
	\begin{center}
		\includegraphics[width=14.6cm]{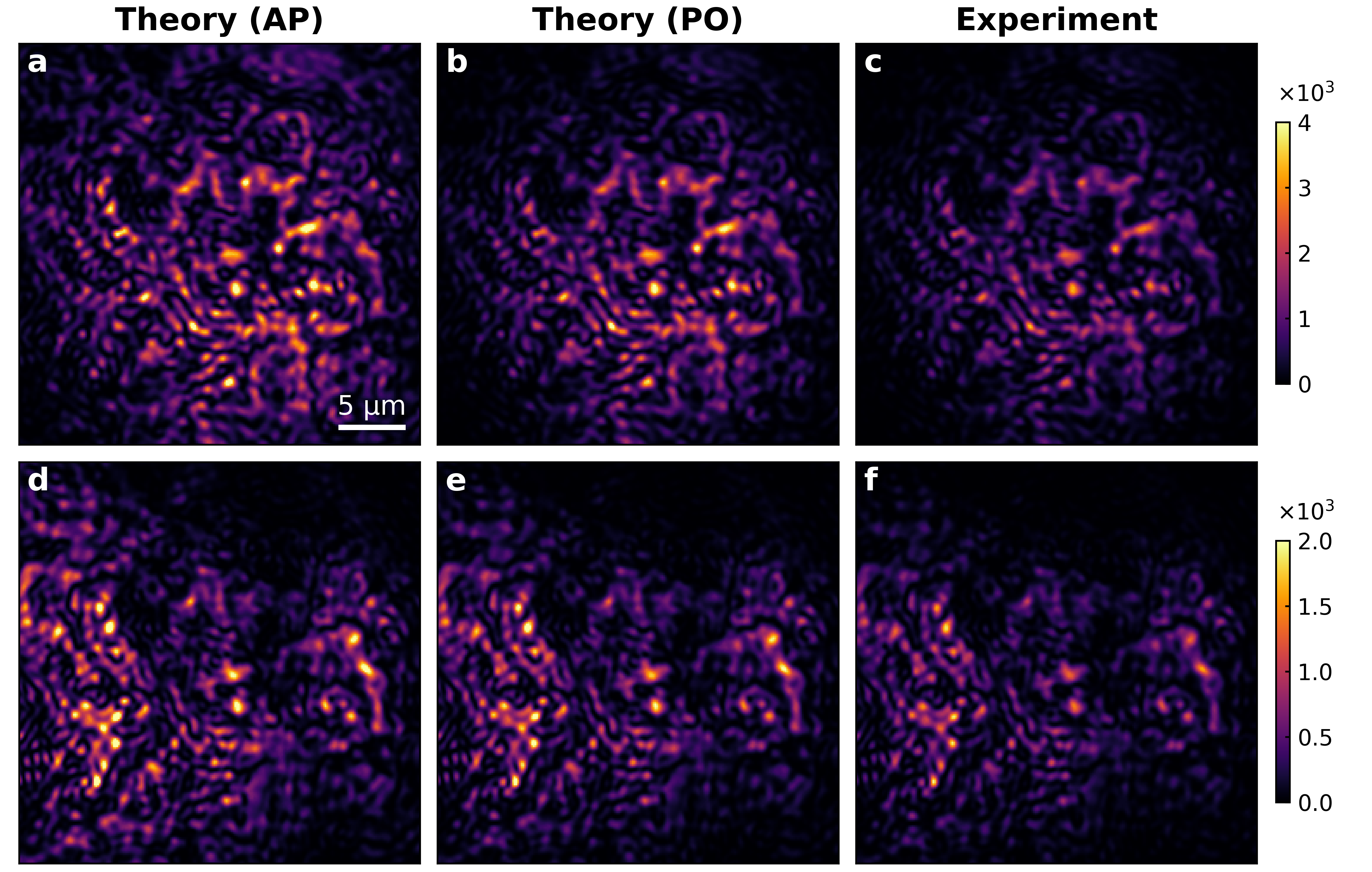}
	\end{center}
	\caption{\textbf{Supplementary Figure S5 | Predicted and measured distributions of the Fisher information per unit area for maximum information states.}
	\textbf{a}, \textbf{b},~Predicted spatial distribution of the normalized Fisher information per unit area for amplitude and phase (AP) modulation and phase-only (PO) modulation, when the observable parameter is the phase shift of the cross-shaped object. \textbf{c},~Measured spatial distribution of the normalized Fisher information per unit area. \textbf{d}--\textbf{f} Analogous to \textbf{a}--\textbf{c} when the observable parameter is the lateral displacement of the circular object.}
\end{figure}

\subsection*{S3.3 -- Correlation between the outgoing field state and its derivative}

In order to characterize the correlation between $|E^\out\ket$ and $|\partial_\theta E^\out\ket$, we calculate the following complex correlation coefficient:
\begin{equation}
\mathscr{C}_\theta=\frac{\bra E^\out | \partial_\theta E^\out \ket }{\Vert E^\out \Vert \cdot \Vert \partial_\theta E^\out \Vert} \; .
\end{equation}
This correlation coefficient is a relevant quantify to assess whether the Fisher information is enclosed in variations of the global phase of the outgoing state as expected for a unitary $S$-matrix ($\mathscr{C}_\theta\simeq \pm i$), or whether it rather enclosed in the state's intensity variations ($\mathscr{C}_\theta\simeq \pm 1$) or speckle decorrelation ($\mathscr{C}_\theta\simeq 0$).

We first calculate the correlation coefficient for each plane wave used to construct the reflection matrix, when the observable parameter is the phase shift $\varphi$ induced by the cross-shaped object (Fig.~S6a) and when it is the lateral position $x$ of the circular object (Fig.~S6b). On average, the measure correlation coefficients are equal to $\mathscr{C}_\varphi=-0.01+0.16\;i$ and $\mathscr{C}_x=0.00+0.01\;i$, respectively. For the maximum information states, the measured correlation coefficients reach $\mathscr{C}_\varphi=0.01+0.96 \, i$ and $\mathscr{C}_x=0.18+0.74 \, i$, respectively. Thus, in both experiments, we observe that the average correlation coefficient for plane wave is close to zero, while the correlation coefficient for the maximum information state is close to the imaginary unit. This observation is likely to reflect the invariance property of maximum information states in the limit of a unitary $S$-matrix, in the same way as the principal modes of a multimode fiber are invariant (to first order) in their output profile with respect to parameter variations except for a global phase shift~\cite{fan_principal_2005_1,carpenter_observation_2015_1,xiong_spatiotemporal_2016_1}. However, in our case, the measured reflection matrices are not unitary, thereby explaining the deviations from this property that are observed in the experiments.

\begin{figure}[ht]
	\begin{center}
		\includegraphics[width=14.6cm]{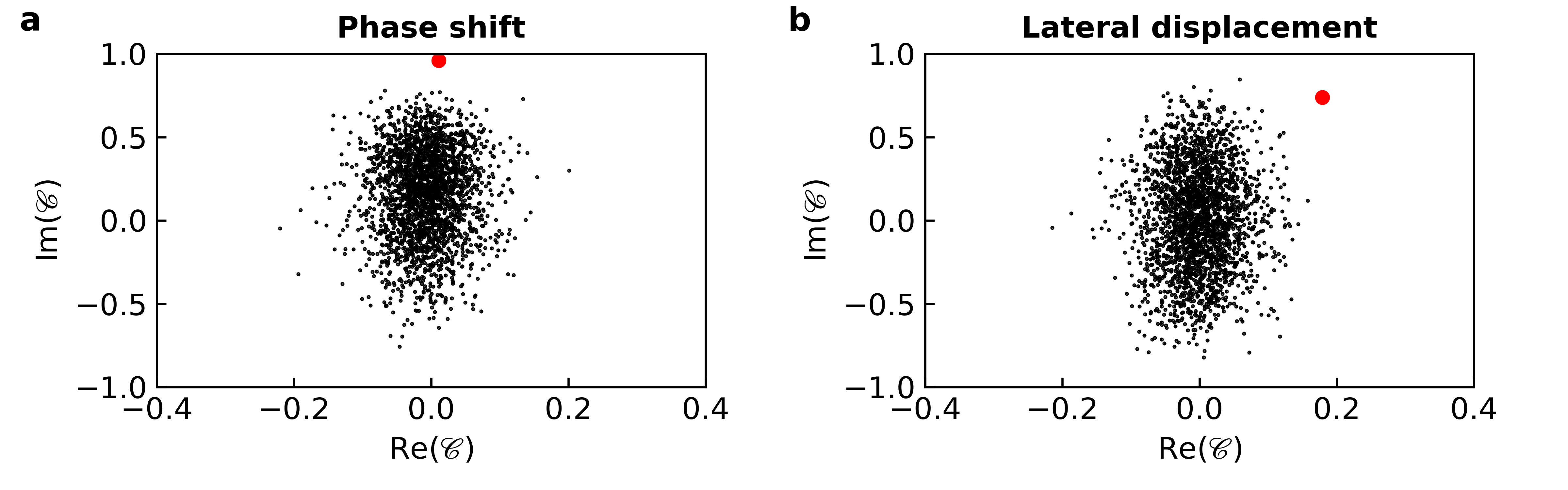}
	\end{center}
	\caption{\textbf{Supplementary Figure S6 | Correlation coefficient between the outgoing field and its derivative.}
	\textbf{a}, Complex correlation coefficient between the outgoing field and its derivative when the observable parameter is the phase shift induced by the cross-shaped object. Black points represent values of the complex correlation coefficient for the $2437$ plane waves used to construct the reflection matrix, and the red point represent the value of the complex correlation coefficient for the optimal incident state. \textbf{b}, Analogous to \textbf{a} when the observable parameter is the lateral displacement of the circular object.}
\end{figure}


\begin{thebibliography}{34}%
	\makeatletter
	\providecommand \@ifxundefined [1]{%
		\@ifx{#1\undefined}
	}%
	\providecommand \@ifnum [1]{%
		\ifnum #1\expandafter \@firstoftwo
		\else \expandafter \@secondoftwo
		\fi
	}%
	\providecommand \@ifx [1]{%
		\ifx #1\expandafter \@firstoftwo
		\else \expandafter \@secondoftwo
		\fi
	}%
	\providecommand \natexlab [1]{#1}%
	\providecommand \enquote  [1]{``#1''}%
	\providecommand \bibnamefont  [1]{#1}%
	\providecommand \bibfnamefont [1]{#1}%
	\providecommand \citenamefont [1]{#1}%
	\providecommand \href@noop [0]{\@secondoftwo}%
	\providecommand \href [0]{\begingroup \@sanitize@url \@href}%
	\providecommand \@href[1]{\@@startlink{#1}\@@href}%
	\providecommand \@@href[1]{\endgroup#1\@@endlink}%
	\providecommand \@sanitize@url [0]{\catcode `\\12\catcode `\$12\catcode
		`\&12\catcode `\#12\catcode `\^12\catcode `\_12\catcode `\%12\relax}%
	\providecommand \@@startlink[1]{}%
	\providecommand \@@endlink[0]{}%
	\providecommand \url  [0]{\begingroup\@sanitize@url \@url }%
	\providecommand \@url [1]{\endgroup\@href {#1}{\urlprefix }}%
	\providecommand \urlprefix  [0]{URL }%
	\providecommand \Eprint [0]{\href }%
	\providecommand \doibase [0]{https://doi.org/}%
	\providecommand \selectlanguage [0]{\@gobble}%
	\providecommand \bibinfo  [0]{\@secondoftwo}%
	\providecommand \bibfield  [0]{\@secondoftwo}%
	\providecommand \translation [1]{[#1]}%
	\providecommand \BibitemOpen [0]{}%
	\providecommand \bibitemStop [0]{}%
	\providecommand \bibitemNoStop [0]{.\EOS\space}%
	\providecommand \EOS [0]{\spacefactor3000\relax}%
	\providecommand \BibitemShut  [1]{\csname bibitem#1\endcsname}%
	\let\auto@bib@innerbib\@empty
	\bibitem [{\citenamefont {Park}\ \emph {et~al.}(2018)\citenamefont {Park},
		\citenamefont {Depeursinge},\ and\ \citenamefont
		{Popescu}}]{park_quantitative_2018}%
	\BibitemOpen
	\bibfield  {author} {\bibinfo {author} {\bibfnamefont {Y.}~\bibnamefont
			{Park}}, \bibinfo {author} {\bibfnamefont {C.}~\bibnamefont {Depeursinge}},\
		and\ \bibinfo {author} {\bibfnamefont {G.}~\bibnamefont {Popescu}},\ }\href
	{https://doi.org/10.1038/s41566-018-0253-x} {\bibfield  {journal} {\bibinfo
			{journal} {Nat. Photonics}\ }\textbf {\bibinfo {volume} {12}},\ \bibinfo
		{pages} {578} (\bibinfo {year} {2018})}\BibitemShut {NoStop}%
	\bibitem [{\citenamefont {Taylor}\ and\ \citenamefont
		{Sandoghdar}(2019)}]{taylor_interferometric_2019}%
	\BibitemOpen
	\bibfield  {author} {\bibinfo {author} {\bibfnamefont {R.~W.}\ \bibnamefont
			{Taylor}}\ and\ \bibinfo {author} {\bibfnamefont {V.}~\bibnamefont
			{Sandoghdar}},\ }\href {https://doi.org/10.1021/acs.nanolett.9b01822}
	{\bibfield  {journal} {\bibinfo  {journal} {Nano Lett.}\ }\textbf {\bibinfo
			{volume} {19}},\ \bibinfo {pages} {4827} (\bibinfo {year}
		{2019})}\BibitemShut {NoStop}%
	\bibitem [{\citenamefont {Osten}\ and\ \citenamefont
		{Reingand}(2012)}]{osten_optical_2012}%
	\BibitemOpen
	\bibfield  {author} {\bibinfo {author} {\bibfnamefont {W.}~\bibnamefont
			{Osten}}\ and\ \bibinfo {author} {\bibfnamefont {N.}~\bibnamefont
			{Reingand}},\ }\href@noop {} {\emph {\bibinfo {title} {Optical {Imaging} and
				{Metrology}: {Advanced} {Technologies}}}}\ (\bibinfo  {publisher} {John Wiley
		\& Sons},\ \bibinfo {year} {2012})\BibitemShut {NoStop}%
	\bibitem [{\citenamefont {van Putten}\ \emph {et~al.}(2012)\citenamefont {van
			Putten}, \citenamefont {Lagendijk},\ and\ \citenamefont
		{Mosk}}]{van_putten_nonimaging_2012}%
	\BibitemOpen
	\bibfield  {author} {\bibinfo {author} {\bibfnamefont {E.~G.}\ \bibnamefont
			{van Putten}}, \bibinfo {author} {\bibfnamefont {A.}~\bibnamefont
			{Lagendijk}},\ and\ \bibinfo {author} {\bibfnamefont {A.~P.}\ \bibnamefont
			{Mosk}},\ }\href {https://doi.org/10.1364/OL.37.001070} {\bibfield  {journal}
		{\bibinfo  {journal} {Opt. Lett.}\ }\textbf {\bibinfo {volume} {37}},\
		\bibinfo {pages} {1070} (\bibinfo {year} {2012})}\BibitemShut {NoStop}%
	\bibitem [{\citenamefont {Shechtman}\ \emph {et~al.}(2014)\citenamefont
		{Shechtman}, \citenamefont {Sahl}, \citenamefont {Backer},\ and\
		\citenamefont {Moerner}}]{shechtman_optimal_2014}%
	\BibitemOpen
	\bibfield  {author} {\bibinfo {author} {\bibfnamefont {Y.}~\bibnamefont
			{Shechtman}}, \bibinfo {author} {\bibfnamefont {S.~J.}\ \bibnamefont {Sahl}},
		\bibinfo {author} {\bibfnamefont {A.~S.}\ \bibnamefont {Backer}},\ and\
		\bibinfo {author} {\bibfnamefont {W.}~\bibnamefont {Moerner}},\ }\href
	{https://doi.org/10.1103/PhysRevLett.113.133902} {\bibfield  {journal}
		{\bibinfo  {journal} {Phys. Rev. Lett.}\ }\textbf {\bibinfo {volume} {113}},\
		\bibinfo {pages} {133902} (\bibinfo {year} {2014})}\BibitemShut {NoStop}%
	\bibitem [{\citenamefont {Balzarotti}\ \emph {et~al.}(2017)\citenamefont
		{Balzarotti}, \citenamefont {Eilers}, \citenamefont {Gwosch}, \citenamefont
		{Gynnå}, \citenamefont {Westphal}, \citenamefont {Stefani}, \citenamefont
		{Elf},\ and\ \citenamefont {Hell}}]{balzarotti_nanometer_2017}%
	\BibitemOpen
	\bibfield  {author} {\bibinfo {author} {\bibfnamefont {F.}~\bibnamefont
			{Balzarotti}}, \bibinfo {author} {\bibfnamefont {Y.}~\bibnamefont {Eilers}},
		\bibinfo {author} {\bibfnamefont {K.~C.}\ \bibnamefont {Gwosch}}, \bibinfo
		{author} {\bibfnamefont {A.~H.}\ \bibnamefont {Gynnå}}, \bibinfo {author}
		{\bibfnamefont {V.}~\bibnamefont {Westphal}}, \bibinfo {author}
		{\bibfnamefont {F.~D.}\ \bibnamefont {Stefani}}, \bibinfo {author}
		{\bibfnamefont {J.}~\bibnamefont {Elf}},\ and\ \bibinfo {author}
		{\bibfnamefont {S.~W.}\ \bibnamefont {Hell}},\ }\href
	{https://doi.org/10.1126/science.aak9913} {\bibfield  {journal} {\bibinfo
			{journal} {Science}\ }\textbf {\bibinfo {volume} {355}},\ \bibinfo {pages}
		{606} (\bibinfo {year} {2017})}\BibitemShut {NoStop}%
	\bibitem [{\citenamefont {Ambichl}\ \emph
		{et~al.}(2017{\natexlab{a}})\citenamefont {Ambichl}, \citenamefont {Xiong},
		\citenamefont {Bromberg}, \citenamefont {Redding}, \citenamefont {Cao},\ and\
		\citenamefont {Rotter}}]{ambichl_super-_2017}%
	\BibitemOpen
	\bibfield  {author} {\bibinfo {author} {\bibfnamefont {P.}~\bibnamefont
			{Ambichl}}, \bibinfo {author} {\bibfnamefont {W.}~\bibnamefont {Xiong}},
		\bibinfo {author} {\bibfnamefont {Y.}~\bibnamefont {Bromberg}}, \bibinfo
		{author} {\bibfnamefont {B.}~\bibnamefont {Redding}}, \bibinfo {author}
		{\bibfnamefont {H.}~\bibnamefont {Cao}},\ and\ \bibinfo {author}
		{\bibfnamefont {S.}~\bibnamefont {Rotter}},\ }\href
	{https://doi.org/10.1103/PhysRevX.7.041053} {\bibfield  {journal} {\bibinfo
			{journal} {Phys. Rev. X}\ }\textbf {\bibinfo {volume} {7}},\ \bibinfo {pages}
		{041053} (\bibinfo {year} {2017}{\natexlab{a}})}\BibitemShut {NoStop}%
	\bibitem [{\citenamefont {Yuan}\ and\ \citenamefont
		{Zheludev}(2019)}]{yuan_detecting_2019}%
	\BibitemOpen
	\bibfield  {author} {\bibinfo {author} {\bibfnamefont {G.~H.}\ \bibnamefont
			{Yuan}}\ and\ \bibinfo {author} {\bibfnamefont {N.~I.}\ \bibnamefont
			{Zheludev}},\ }\href {https://doi.org/10.1126/science.aaw7840} {\bibfield
		{journal} {\bibinfo  {journal} {Science}\ }\textbf {\bibinfo {volume}
			{364}},\ \bibinfo {pages} {771} (\bibinfo {year} {2019})}\BibitemShut
	{NoStop}%
	\bibitem [{\citenamefont {Juffmann}\ \emph {et~al.}(2020)\citenamefont
		{Juffmann}, \citenamefont {de~los Ríos~Sommer},\ and\ \citenamefont
		{Gigan}}]{juffmann_local_2020}%
	\BibitemOpen
	\bibfield  {author} {\bibinfo {author} {\bibfnamefont {T.}~\bibnamefont
			{Juffmann}}, \bibinfo {author} {\bibfnamefont {A.}~\bibnamefont {de~los
				Ríos~Sommer}},\ and\ \bibinfo {author} {\bibfnamefont {S.}~\bibnamefont
			{Gigan}},\ }\href {https://doi.org/10.1016/j.optcom.2019.124484} {\bibfield
		{journal} {\bibinfo  {journal} {Opt. Commun.}\ }\textbf {\bibinfo {volume}
			{454}},\ \bibinfo {pages} {124484} (\bibinfo {year} {2020})}\BibitemShut
	{NoStop}%
	\bibitem [{\citenamefont {Bouchet}\ \emph {et~al.}(2020)\citenamefont
		{Bouchet}, \citenamefont {Carminati},\ and\ \citenamefont
		{Mosk}}]{bouchet_influence_2020}%
	\BibitemOpen
	\bibfield  {author} {\bibinfo {author} {\bibfnamefont {D.}~\bibnamefont
			{Bouchet}}, \bibinfo {author} {\bibfnamefont {R.}~\bibnamefont {Carminati}},\
		and\ \bibinfo {author} {\bibfnamefont {A.~P.}\ \bibnamefont {Mosk}},\ }\href
	{https://doi.org/10.1103/PhysRevLett.124.133903} {\bibfield  {journal}
		{\bibinfo  {journal} {Phys. Rev. Lett.}\ }\textbf {\bibinfo {volume} {124}},\
		\bibinfo {pages} {133903} (\bibinfo {year} {2020})}\BibitemShut {NoStop}%
	\bibitem [{\citenamefont {Giovannetti}\ \emph {et~al.}(2011)\citenamefont
		{Giovannetti}, \citenamefont {Lloyd},\ and\ \citenamefont
		{Maccone}}]{giovannetti_advances_2011}%
	\BibitemOpen
	\bibfield  {author} {\bibinfo {author} {\bibfnamefont {V.}~\bibnamefont
			{Giovannetti}}, \bibinfo {author} {\bibfnamefont {S.}~\bibnamefont {Lloyd}},\
		and\ \bibinfo {author} {\bibfnamefont {L.}~\bibnamefont {Maccone}},\ }\href
	{https://doi.org/10.1038/nphoton.2011.35} {\bibfield  {journal} {\bibinfo
			{journal} {Nat. Photonics}\ }\textbf {\bibinfo {volume} {5}},\ \bibinfo
		{pages} {222} (\bibinfo {year} {2011})}\BibitemShut {NoStop}%
	\bibitem [{\citenamefont {Rotter}\ and\ \citenamefont
		{Gigan}(2017)}]{rotter_light_2017}%
	\BibitemOpen
	\bibfield  {author} {\bibinfo {author} {\bibfnamefont {S.}~\bibnamefont
			{Rotter}}\ and\ \bibinfo {author} {\bibfnamefont {S.}~\bibnamefont {Gigan}},\
	}\href {https://doi.org/10.1103/RevModPhys.89.015005} {\bibfield  {journal}
		{\bibinfo  {journal} {Rev. Mod. Phys.}\ }\textbf {\bibinfo {volume} {89}},\
		\bibinfo {pages} {015005} (\bibinfo {year} {2017})}\BibitemShut {NoStop}%
	\bibitem [{\citenamefont {Barrett}\ and\ \citenamefont
		{Myers}(2013)}]{barrett_foundations_2013}%
	\BibitemOpen
	\bibfield  {author} {\bibinfo {author} {\bibfnamefont {H.~H.}\ \bibnamefont
			{Barrett}}\ and\ \bibinfo {author} {\bibfnamefont {K.~J.}\ \bibnamefont
			{Myers}},\ }\href@noop {} {\emph {\bibinfo {title} {Foundations of {Image}
				{Science}}}}\ (\bibinfo  {publisher} {John Wiley \& Sons},\ \bibinfo {year}
	{2013})\BibitemShut {NoStop}%
	\bibitem [{\citenamefont {Kay}(1993)}]{kay_fundamentals_1993}%
	\BibitemOpen
	\bibfield  {author} {\bibinfo {author} {\bibfnamefont {S.}~\bibnamefont
			{Kay}},\ }\href@noop {} {\emph {\bibinfo {title} {Fundamentals of
				{Statistical} {Processing}, {Volume} {I}: {Estimation} {Theory}}}}\ (\bibinfo
	{publisher} {Prentice Hall},\ \bibinfo {year} {1993})\BibitemShut {NoStop}%
	\bibitem [{\citenamefont {Ambichl}\ \emph
		{et~al.}(2017{\natexlab{b}})\citenamefont {Ambichl}, \citenamefont
		{Brandstötter}, \citenamefont {Böhm}, \citenamefont {Kühmayer},
		\citenamefont {Kuhl},\ and\ \citenamefont {Rotter}}]{ambichl_focusing_2017}%
	\BibitemOpen
	\bibfield  {author} {\bibinfo {author} {\bibfnamefont {P.}~\bibnamefont
			{Ambichl}}, \bibinfo {author} {\bibfnamefont {A.}~\bibnamefont
			{Brandstötter}}, \bibinfo {author} {\bibfnamefont {J.}~\bibnamefont
			{Böhm}}, \bibinfo {author} {\bibfnamefont {M.}~\bibnamefont {Kühmayer}},
		\bibinfo {author} {\bibfnamefont {U.}~\bibnamefont {Kuhl}},\ and\ \bibinfo
		{author} {\bibfnamefont {S.}~\bibnamefont {Rotter}},\ }\href
	{https://doi.org/10.1103/PhysRevLett.119.033903} {\bibfield  {journal}
		{\bibinfo  {journal} {Phys. Rev. Lett.}\ }\textbf {\bibinfo {volume} {119}},\
		\bibinfo {pages} {033903} (\bibinfo {year} {2017}{\natexlab{b}})}\BibitemShut
	{NoStop}%
	\bibitem [{\citenamefont {Horodynski}\ \emph {et~al.}(2020)\citenamefont
		{Horodynski}, \citenamefont {Kühmayer}, \citenamefont {Brandstötter},
		\citenamefont {Pichler}, \citenamefont {Fyodorov}, \citenamefont {Kuhl},\
		and\ \citenamefont {Rotter}}]{horodynski_optimal_2020}%
	\BibitemOpen
	\bibfield  {author} {\bibinfo {author} {\bibfnamefont {M.}~\bibnamefont
			{Horodynski}}, \bibinfo {author} {\bibfnamefont {M.}~\bibnamefont
			{Kühmayer}}, \bibinfo {author} {\bibfnamefont {A.}~\bibnamefont
			{Brandstötter}}, \bibinfo {author} {\bibfnamefont {K.}~\bibnamefont
			{Pichler}}, \bibinfo {author} {\bibfnamefont {Y.~V.}\ \bibnamefont
			{Fyodorov}}, \bibinfo {author} {\bibfnamefont {U.}~\bibnamefont {Kuhl}},\
		and\ \bibinfo {author} {\bibfnamefont {S.}~\bibnamefont {Rotter}},\ }\href
	{https://doi.org/10.1038/s41566-019-0550-z} {\bibfield  {journal} {\bibinfo
			{journal} {Nat. Photonics}\ }\textbf {\bibinfo {volume} {14}},\ \bibinfo
		{pages} {149} (\bibinfo {year} {2020})}\BibitemShut {NoStop}%
	\bibitem [{\citenamefont {Mosk}\ \emph {et~al.}(2012)\citenamefont {Mosk},
		\citenamefont {Lagendijk}, \citenamefont {Lerosey},\ and\ \citenamefont
		{Fink}}]{mosk_controlling_2012}%
	\BibitemOpen
	\bibfield  {author} {\bibinfo {author} {\bibfnamefont {A.~P.}\ \bibnamefont
			{Mosk}}, \bibinfo {author} {\bibfnamefont {A.}~\bibnamefont {Lagendijk}},
		\bibinfo {author} {\bibfnamefont {G.}~\bibnamefont {Lerosey}},\ and\ \bibinfo
		{author} {\bibfnamefont {M.}~\bibnamefont {Fink}},\ }\href
	{https://doi.org/10.1038/nphoton.2012.88} {\bibfield  {journal} {\bibinfo
			{journal} {Nat. Photonics}\ }\textbf {\bibinfo {volume} {6}},\ \bibinfo
		{pages} {283} (\bibinfo {year} {2012})}\BibitemShut {NoStop}%
	\bibitem [{\citenamefont {Fang}\ \emph {et~al.}(2018)\citenamefont {Fang},
		\citenamefont {Zhao},\ and\ \citenamefont
		{Tian}}]{fang_concentration--measure_2018}%
	\BibitemOpen
	\bibfield  {author} {\bibinfo {author} {\bibfnamefont {P.}~\bibnamefont
			{Fang}}, \bibinfo {author} {\bibfnamefont {L.}~\bibnamefont {Zhao}},\ and\
		\bibinfo {author} {\bibfnamefont {C.}~\bibnamefont {Tian}},\ }\href
	{https://doi.org/10.1103/PhysRevLett.121.140603} {\bibfield  {journal}
		{\bibinfo  {journal} {Phys. Rev. Lett.}\ }\textbf {\bibinfo {volume} {121}},\
		\bibinfo {pages} {140603} (\bibinfo {year} {2018})}\BibitemShut {NoStop}%
	\bibitem [{\citenamefont {Braunstein}\ \emph {et~al.}(1996)\citenamefont
		{Braunstein}, \citenamefont {Caves},\ and\ \citenamefont
		{Milburn}}]{braunstein_generalized_1996}%
	\BibitemOpen
	\bibfield  {author} {\bibinfo {author} {\bibfnamefont {S.~L.}\ \bibnamefont
			{Braunstein}}, \bibinfo {author} {\bibfnamefont {C.~M.}\ \bibnamefont
			{Caves}},\ and\ \bibinfo {author} {\bibfnamefont {G.~J.}\ \bibnamefont
			{Milburn}},\ }\href {https://doi.org/10.1006/aphy.1996.0040} {\bibfield
		{journal} {\bibinfo  {journal} {Ann. Phys. (N. Y.)}\ }\textbf {\bibinfo
			{volume} {247}},\ \bibinfo {pages} {135} (\bibinfo {year}
		{1996})}\BibitemShut {NoStop}%
	\bibitem [{\citenamefont {Zhou}\ \emph {et~al.}(2014)\citenamefont {Zhou},
		\citenamefont {Ruan}, \citenamefont {Yang},\ and\ \citenamefont
		{Judkewitz}}]{zhou_focusing_2014}%
	\BibitemOpen
	\bibfield  {author} {\bibinfo {author} {\bibfnamefont {E.~H.}\ \bibnamefont
			{Zhou}}, \bibinfo {author} {\bibfnamefont {H.}~\bibnamefont {Ruan}}, \bibinfo
		{author} {\bibfnamefont {C.}~\bibnamefont {Yang}},\ and\ \bibinfo {author}
		{\bibfnamefont {B.}~\bibnamefont {Judkewitz}},\ }\href
	{https://doi.org/10.1364/OPTICA.1.000227} {\bibfield  {journal} {\bibinfo
			{journal} {Optica}\ }\textbf {\bibinfo {volume} {1}},\ \bibinfo {pages} {227}
		(\bibinfo {year} {2014})}\BibitemShut {NoStop}%
	\bibitem [{\citenamefont {Ma}\ \emph {et~al.}(2014)\citenamefont {Ma},
		\citenamefont {Xu}, \citenamefont {Liu},\ and\ \citenamefont
		{Wang}}]{ma_time-reversed_2014}%
	\BibitemOpen
	\bibfield  {author} {\bibinfo {author} {\bibfnamefont {C.}~\bibnamefont
			{Ma}}, \bibinfo {author} {\bibfnamefont {X.}~\bibnamefont {Xu}}, \bibinfo
		{author} {\bibfnamefont {Y.}~\bibnamefont {Liu}},\ and\ \bibinfo {author}
		{\bibfnamefont {L.~V.}\ \bibnamefont {Wang}},\ }\href
	{https://doi.org/10.1038/nphoton.2014.251} {\bibfield  {journal} {\bibinfo
			{journal} {Nat. Photonics}\ }\textbf {\bibinfo {volume} {8}},\ \bibinfo
		{pages} {931} (\bibinfo {year} {2014})}\BibitemShut {NoStop}%
	\bibitem [{\citenamefont {Ruan}\ \emph {et~al.}(2017)\citenamefont {Ruan},
		\citenamefont {Haber}, \citenamefont {Liu}, \citenamefont {Brake},
		\citenamefont {Kim}, \citenamefont {Berlin},\ and\ \citenamefont
		{Yang}}]{ruan_focusing_2017}%
	\BibitemOpen
	\bibfield  {author} {\bibinfo {author} {\bibfnamefont {H.}~\bibnamefont
			{Ruan}}, \bibinfo {author} {\bibfnamefont {T.}~\bibnamefont {Haber}},
		\bibinfo {author} {\bibfnamefont {Y.}~\bibnamefont {Liu}}, \bibinfo {author}
		{\bibfnamefont {J.}~\bibnamefont {Brake}}, \bibinfo {author} {\bibfnamefont
			{J.}~\bibnamefont {Kim}}, \bibinfo {author} {\bibfnamefont {J.~M.}\
			\bibnamefont {Berlin}},\ and\ \bibinfo {author} {\bibfnamefont
			{C.}~\bibnamefont {Yang}},\ }\href {https://doi.org/10.1364/OPTICA.4.001337}
	{\bibfield  {journal} {\bibinfo  {journal} {Optica}\ }\textbf {\bibinfo
			{volume} {4}},\ \bibinfo {pages} {1337} (\bibinfo {year} {2017})}\BibitemShut
	{NoStop}%
	\bibitem [{\citenamefont {Carpenter}\ \emph {et~al.}(2015)\citenamefont
		{Carpenter}, \citenamefont {Eggleton},\ and\ \citenamefont
		{Schröder}}]{carpenter_observation_2015}%
	\BibitemOpen
	\bibfield  {author} {\bibinfo {author} {\bibfnamefont {J.}~\bibnamefont
			{Carpenter}}, \bibinfo {author} {\bibfnamefont {B.~J.}\ \bibnamefont
			{Eggleton}},\ and\ \bibinfo {author} {\bibfnamefont {J.}~\bibnamefont
			{Schröder}},\ }\href {https://doi.org/10.1038/nphoton.2015.188} {\bibfield
		{journal} {\bibinfo  {journal} {Nat. Photonics}\ }\textbf {\bibinfo {volume}
			{9}},\ \bibinfo {pages} {751} (\bibinfo {year} {2015})}\BibitemShut {NoStop}%
	\bibitem [{\citenamefont {Xiong}\ \emph {et~al.}(2016)\citenamefont {Xiong},
		\citenamefont {Ambichl}, \citenamefont {Bromberg}, \citenamefont {Redding},
		\citenamefont {Rotter},\ and\ \citenamefont
		{Cao}}]{xiong_spatiotemporal_2016}%
	\BibitemOpen
	\bibfield  {author} {\bibinfo {author} {\bibfnamefont {W.}~\bibnamefont
			{Xiong}}, \bibinfo {author} {\bibfnamefont {P.}~\bibnamefont {Ambichl}},
		\bibinfo {author} {\bibfnamefont {Y.}~\bibnamefont {Bromberg}}, \bibinfo
		{author} {\bibfnamefont {B.}~\bibnamefont {Redding}}, \bibinfo {author}
		{\bibfnamefont {S.}~\bibnamefont {Rotter}},\ and\ \bibinfo {author}
		{\bibfnamefont {H.}~\bibnamefont {Cao}},\ }\href
	{https://doi.org/10.1103/PhysRevLett.117.053901} {\bibfield  {journal}
		{\bibinfo  {journal} {Phys. Rev. Lett.}\ }\textbf {\bibinfo {volume} {117}},\
		\bibinfo {pages} {053901} (\bibinfo {year} {2016})}\BibitemShut {NoStop}%
	\bibitem [{\citenamefont {Hodaei}\ \emph {et~al.}(2017)\citenamefont {Hodaei},
		\citenamefont {Hassan}, \citenamefont {Wittek}, \citenamefont
		{Garcia-Gracia}, \citenamefont {El-Ganainy}, \citenamefont
		{Christodoulides},\ and\ \citenamefont {Khajavikhan}}]{hodaei_enhanced_2017}%
	\BibitemOpen
	\bibfield  {author} {\bibinfo {author} {\bibfnamefont {H.}~\bibnamefont
			{Hodaei}}, \bibinfo {author} {\bibfnamefont {A.~U.}\ \bibnamefont {Hassan}},
		\bibinfo {author} {\bibfnamefont {S.}~\bibnamefont {Wittek}}, \bibinfo
		{author} {\bibfnamefont {H.}~\bibnamefont {Garcia-Gracia}}, \bibinfo {author}
		{\bibfnamefont {R.}~\bibnamefont {El-Ganainy}}, \bibinfo {author}
		{\bibfnamefont {D.~N.}\ \bibnamefont {Christodoulides}},\ and\ \bibinfo
		{author} {\bibfnamefont {M.}~\bibnamefont {Khajavikhan}},\ }\href
	{https://doi.org/10.1038/nature23280} {\bibfield  {journal} {\bibinfo
			{journal} {Nature}\ }\textbf {\bibinfo {volume} {548}},\ \bibinfo {pages}
		{187} (\bibinfo {year} {2017})}\BibitemShut {NoStop}%
	\bibitem [{\citenamefont {Chen}\ \emph {et~al.}(2017)\citenamefont {Chen},
		\citenamefont {Kaya~Özdemir}, \citenamefont {Zhao}, \citenamefont
		{Wiersig},\ and\ \citenamefont {Yang}}]{chen_exceptional_2017}%
	\BibitemOpen
	\bibfield  {author} {\bibinfo {author} {\bibfnamefont {W.}~\bibnamefont
			{Chen}}, \bibinfo {author} {\bibfnamefont {S.}~\bibnamefont {Kaya~Özdemir}},
		\bibinfo {author} {\bibfnamefont {G.}~\bibnamefont {Zhao}}, \bibinfo {author}
		{\bibfnamefont {J.}~\bibnamefont {Wiersig}},\ and\ \bibinfo {author}
		{\bibfnamefont {L.}~\bibnamefont {Yang}},\ }\href
	{https://doi.org/10.1038/nature23281} {\bibfield  {journal} {\bibinfo
			{journal} {Nature}\ }\textbf {\bibinfo {volume} {548}},\ \bibinfo {pages}
		{192} (\bibinfo {year} {2017})}\BibitemShut {NoStop}%
	\bibitem [{\citenamefont {Gérardin}\ \emph {et~al.}(2014)\citenamefont
		{Gérardin}, \citenamefont {Laurent}, \citenamefont {Derode}, \citenamefont
		{Prada},\ and\ \citenamefont {Aubry}}]{gerardin_full_2014}%
	\BibitemOpen
	\bibfield  {author} {\bibinfo {author} {\bibfnamefont {B.}~\bibnamefont
			{Gérardin}}, \bibinfo {author} {\bibfnamefont {J.}~\bibnamefont {Laurent}},
		\bibinfo {author} {\bibfnamefont {A.}~\bibnamefont {Derode}}, \bibinfo
		{author} {\bibfnamefont {C.}~\bibnamefont {Prada}},\ and\ \bibinfo {author}
		{\bibfnamefont {A.}~\bibnamefont {Aubry}},\ }\href
	{https://doi.org/10.1103/PhysRevLett.113.173901} {\bibfield  {journal}
		{\bibinfo  {journal} {Phys. Rev. Lett.}\ }\textbf {\bibinfo {volume} {113}},\
		\bibinfo {pages} {173901} (\bibinfo {year} {2014})}\BibitemShut {NoStop}%
	\bibitem [{\citenamefont {Shi}\ and\ \citenamefont
		{Genack}(2012)}]{shi_transmission_2012}%
	\BibitemOpen
	\bibfield  {author} {\bibinfo {author} {\bibfnamefont {Z.}~\bibnamefont
			{Shi}}\ and\ \bibinfo {author} {\bibfnamefont {A.~Z.}\ \bibnamefont
			{Genack}},\ }\href {https://doi.org/10.1103/PhysRevLett.108.043901}
	{\bibfield  {journal} {\bibinfo  {journal} {Phys. Rev. Lett.}\ }\textbf
		{\bibinfo {volume} {108}},\ \bibinfo {pages} {043901} (\bibinfo {year}
		{2012})}\BibitemShut {NoStop}%
	\bibitem [{\citenamefont {Giovannetti}\ \emph {et~al.}(2006)\citenamefont
		{Giovannetti}, \citenamefont {Lloyd},\ and\ \citenamefont
		{Maccone}}]{giovannetti_quantum_2006}%
	\BibitemOpen
	\bibfield  {author} {\bibinfo {author} {\bibfnamefont {V.}~\bibnamefont
			{Giovannetti}}, \bibinfo {author} {\bibfnamefont {S.}~\bibnamefont {Lloyd}},\
		and\ \bibinfo {author} {\bibfnamefont {L.}~\bibnamefont {Maccone}},\ }\href
	{https://doi.org/10.1103/PhysRevLett.96.010401} {\bibfield  {journal}
		{\bibinfo  {journal} {Phys. Rev. Lett.}\ }\textbf {\bibinfo {volume} {96}},\
		\bibinfo {pages} {010401} (\bibinfo {year} {2006})}\BibitemShut {NoStop}%
	\bibitem [{\citenamefont {Fiderer}\ \emph {et~al.}(2019)\citenamefont
		{Fiderer}, \citenamefont {Fraïsse},\ and\ \citenamefont
		{Braun}}]{fiderer_maximal_2019}%
	\BibitemOpen
	\bibfield  {author} {\bibinfo {author} {\bibfnamefont {L.~J.}\ \bibnamefont
			{Fiderer}}, \bibinfo {author} {\bibfnamefont {J.~M.}\ \bibnamefont
			{Fraïsse}},\ and\ \bibinfo {author} {\bibfnamefont {D.}~\bibnamefont
			{Braun}},\ }\href {https://doi.org/10.1103/PhysRevLett.123.250502} {\bibfield
		{journal} {\bibinfo  {journal} {Phys. Rev. Lett.}\ }\textbf {\bibinfo
			{volume} {123}},\ \bibinfo {pages} {250502} (\bibinfo {year}
		{2019})}\BibitemShut {NoStop}%
	\bibitem [{\citenamefont {Popoff}\ \emph {et~al.}(2010)\citenamefont {Popoff},
		\citenamefont {Lerosey}, \citenamefont {Carminati}, \citenamefont {Fink},
		\citenamefont {Boccara},\ and\ \citenamefont
		{Gigan}}]{popoff_measuring_2010}%
	\BibitemOpen
	\bibfield  {author} {\bibinfo {author} {\bibfnamefont {S.~M.}\ \bibnamefont
			{Popoff}}, \bibinfo {author} {\bibfnamefont {G.}~\bibnamefont {Lerosey}},
		\bibinfo {author} {\bibfnamefont {R.}~\bibnamefont {Carminati}}, \bibinfo
		{author} {\bibfnamefont {M.}~\bibnamefont {Fink}}, \bibinfo {author}
		{\bibfnamefont {A.~C.}\ \bibnamefont {Boccara}},\ and\ \bibinfo {author}
		{\bibfnamefont {S.}~\bibnamefont {Gigan}},\ }\href
	{https://doi.org/10.1103/PhysRevLett.104.100601} {\bibfield  {journal}
		{\bibinfo  {journal} {Phys. Rev. Lett.}\ }\textbf {\bibinfo {volume} {104}},\
		\bibinfo {pages} {100601} (\bibinfo {year} {2010})}\BibitemShut {NoStop}%
	\bibitem [{\citenamefont {Pai}\ \emph {et~al.}(2020)\citenamefont {Pai},
		\citenamefont {Bosch},\ and\ \citenamefont {Mosk}}]{pai_optical_2020}%
	\BibitemOpen
	\bibfield  {author} {\bibinfo {author} {\bibfnamefont {P.}~\bibnamefont
			{Pai}}, \bibinfo {author} {\bibfnamefont {J.}~\bibnamefont {Bosch}},\ and\
		\bibinfo {author} {\bibfnamefont {A.~P.}\ \bibnamefont {Mosk}},\ }\href
	{https://doi.org/10.1364/OSAC.384832} {\bibfield  {journal} {\bibinfo
			{journal} {OSA Continuum}\ }\textbf {\bibinfo {volume} {3}},\ \bibinfo
		{pages} {637} (\bibinfo {year} {2020})}\BibitemShut {NoStop}%
	\bibitem [{\citenamefont {Takeda}\ \emph {et~al.}(1982)\citenamefont {Takeda},
		\citenamefont {Ina},\ and\ \citenamefont
		{Kobayashi}}]{takeda_fourier-transform_1982}%
	\BibitemOpen
	\bibfield  {author} {\bibinfo {author} {\bibfnamefont {M.}~\bibnamefont
			{Takeda}}, \bibinfo {author} {\bibfnamefont {H.}~\bibnamefont {Ina}},\ and\
		\bibinfo {author} {\bibfnamefont {S.}~\bibnamefont {Kobayashi}},\ }\href
	{https://doi.org/10.1364/JOSA.72.000156} {\bibfield  {journal} {\bibinfo
			{journal} {J. Opt. Soc. Am.}\ }\textbf {\bibinfo {volume} {72}},\ \bibinfo
		{pages} {156} (\bibinfo {year} {1982})}\BibitemShut {NoStop}%
	\bibitem [{\citenamefont {Cuche}\ \emph {et~al.}(2000)\citenamefont {Cuche},
		\citenamefont {Marquet},\ and\ \citenamefont
		{Depeursinge}}]{cuche_spatial_2000}%
	\BibitemOpen
	\bibfield  {author} {\bibinfo {author} {\bibfnamefont {E.}~\bibnamefont
			{Cuche}}, \bibinfo {author} {\bibfnamefont {P.}~\bibnamefont {Marquet}},\
		and\ \bibinfo {author} {\bibfnamefont {C.}~\bibnamefont {Depeursinge}},\
	}\href {https://doi.org/10.1364/AO.39.004070} {\bibfield  {journal} {\bibinfo
			{journal} {Appl. Opt.}\ }\textbf {\bibinfo {volume} {39}},\ \bibinfo {pages}
		{4070} (\bibinfo {year} {2000})}\BibitemShut {NoStop}%
\end{thebibliography}

\begin{thebibliography}{16}%
	\makeatletter
	\providecommand \@ifxundefined [1]{%
		\@ifx{#1\undefined}
	}%
	\providecommand \@ifnum [1]{%
		\ifnum #1\expandafter \@firstoftwo
		\else \expandafter \@secondoftwo
		\fi
	}%
	\providecommand \@ifx [1]{%
		\ifx #1\expandafter \@firstoftwo
		\else \expandafter \@secondoftwo
		\fi
	}%
	\providecommand \natexlab [1]{#1}%
	\providecommand \enquote  [1]{``#1''}%
	\providecommand \bibnamefont  [1]{#1}%
	\providecommand \bibfnamefont [1]{#1}%
	\providecommand \citenamefont [1]{#1}%
	\providecommand \href@noop [0]{\@secondoftwo}%
	\providecommand \href [0]{\begingroup \@sanitize@url \@href}%
	\providecommand \@href[1]{\@@startlink{#1}\@@href}%
	\providecommand \@@href[1]{\endgroup#1\@@endlink}%
	\providecommand \@sanitize@url [0]{\catcode `\\12\catcode `\$12\catcode
		`\&12\catcode `\#12\catcode `\^12\catcode `\_12\catcode `\%12\relax}%
	\providecommand \@@startlink[1]{}%
	\providecommand \@@endlink[0]{}%
	\providecommand \url  [0]{\begingroup\@sanitize@url \@url }%
	\providecommand \@url [1]{\endgroup\@href {#1}{\urlprefix }}%
	\providecommand \urlprefix  [0]{URL }%
	\providecommand \Eprint [0]{\href }%
	\providecommand \doibase [0]{https://doi.org/}%
	\providecommand \selectlanguage [0]{\@gobble}%
	\providecommand \bibinfo  [0]{\@secondoftwo}%
	\providecommand \bibfield  [0]{\@secondoftwo}%
	\providecommand \translation [1]{[#1]}%
	\providecommand \BibitemOpen [0]{}%
	\providecommand \bibitemStop [0]{}%
	\providecommand \bibitemNoStop [0]{.\EOS\space}%
	\providecommand \EOS [0]{\spacefactor3000\relax}%
	\providecommand \BibitemShut  [1]{\csname bibitem#1\endcsname}%
	\let\auto@bib@innerbib\@empty
	\bibitem [{\citenamefont {Kay}(1993)}]{kay_fundamentals_1993_1}%
	\BibitemOpen
	\bibfield  {author} {\bibinfo {author} {\bibfnamefont {S.}~\bibnamefont
			{Kay}},\ }\href@noop {} {\emph {\bibinfo {title} {Fundamentals of
				{Statistical} {Processing}, {Volume} {I}: {Estimation} {Theory}}}}\ (\bibinfo
	{publisher} {Prentice Hall},\ \bibinfo {year} {1993})\BibitemShut {NoStop}%
	\bibitem [{\citenamefont {Mandel}\ and\ \citenamefont
		{Wolf}(1995)}]{mandel_optical_1995_1}%
	\BibitemOpen
	\bibfield  {author} {\bibinfo {author} {\bibfnamefont {L.}~\bibnamefont
			{Mandel}}\ and\ \bibinfo {author} {\bibfnamefont {E.}~\bibnamefont {Wolf}},\
	}\href@noop {} {\emph {\bibinfo {title} {Optical {Coherence} and {Quantum}
				{Optics}}}}\ (\bibinfo  {publisher} {Cambridge University Press},\ \bibinfo
	{year} {1995})\BibitemShut {NoStop}%
	\bibitem [{\citenamefont {Helstrom}(1969)}]{helstrom_quantum_1969_1}%
	\BibitemOpen
	\bibfield  {author} {\bibinfo {author} {\bibfnamefont {C.~W.}\ \bibnamefont
			{Helstrom}},\ }\href {https://doi.org/10.1007/BF01007479} {\bibfield
		{journal} {\bibinfo  {journal} {J. Stat. Phys.}\ }\textbf {\bibinfo {volume}
			{1}},\ \bibinfo {pages} {231} (\bibinfo {year} {1969})}\BibitemShut {NoStop}%
	\bibitem [{\citenamefont {Braunstein}\ \emph {et~al.}(1996)\citenamefont
		{Braunstein}, \citenamefont {Caves},\ and\ \citenamefont
		{Milburn}}]{braunstein_generalized_1996_1}%
	\BibitemOpen
	\bibfield  {author} {\bibinfo {author} {\bibfnamefont {S.~L.}\ \bibnamefont
			{Braunstein}}, \bibinfo {author} {\bibfnamefont {C.~M.}\ \bibnamefont
			{Caves}},\ and\ \bibinfo {author} {\bibfnamefont {G.~J.}\ \bibnamefont
			{Milburn}},\ }\href {https://doi.org/10.1006/aphy.1996.0040} {\bibfield
		{journal} {\bibinfo  {journal} {Ann. Phys. (N. Y.)}\ }\textbf {\bibinfo
			{volume} {247}},\ \bibinfo {pages} {135} (\bibinfo {year}
		{1996})}\BibitemShut {NoStop}%
	\bibitem [{\citenamefont {Demkowicz-Dobrzański}\ \emph
		{et~al.}(2015)\citenamefont {Demkowicz-Dobrzański}, \citenamefont
		{Jarzyna},\ and\ \citenamefont
		{Kołodyński}}]{demkowicz-dobrzanski_chapter_2015_1}%
	\BibitemOpen
	\bibfield  {author} {\bibinfo {author} {\bibfnamefont {R.}~\bibnamefont
			{Demkowicz-Dobrzański}}, \bibinfo {author} {\bibfnamefont {M.}~\bibnamefont
			{Jarzyna}},\ and\ \bibinfo {author} {\bibfnamefont {J.}~\bibnamefont
			{Kołodyński}},\ }in\ \href {https://doi.org/10.1016/bs.po.2015.02.003}
	{\emph {\bibinfo {booktitle} {Progress in {Optics}}}},\ Vol.~\bibinfo
	{volume} {60}\ (\bibinfo  {publisher} {Elsevier},\ \bibinfo {year} {2015})\
	pp.\ \bibinfo {pages} {345--435}\BibitemShut {NoStop}%
	\bibitem [{\citenamefont {Ambichl}\ \emph {et~al.}(2017)\citenamefont
		{Ambichl}, \citenamefont {Brandstötter}, \citenamefont {Böhm},
		\citenamefont {Kühmayer}, \citenamefont {Kuhl},\ and\ \citenamefont
		{Rotter}}]{ambichl_focusing_2017_1}%
	\BibitemOpen
	\bibfield  {author} {\bibinfo {author} {\bibfnamefont {P.}~\bibnamefont
			{Ambichl}}, \bibinfo {author} {\bibfnamefont {A.}~\bibnamefont
			{Brandstötter}}, \bibinfo {author} {\bibfnamefont {J.}~\bibnamefont
			{Böhm}}, \bibinfo {author} {\bibfnamefont {M.}~\bibnamefont {Kühmayer}},
		\bibinfo {author} {\bibfnamefont {U.}~\bibnamefont {Kuhl}},\ and\ \bibinfo
		{author} {\bibfnamefont {S.}~\bibnamefont {Rotter}},\ }\href
	{https://doi.org/10.1103/PhysRevLett.119.033903} {\bibfield  {journal}
		{\bibinfo  {journal} {Phys. Rev. Lett.}\ }\textbf {\bibinfo {volume} {119}},\
		\bibinfo {pages} {033903} (\bibinfo {year} {2017})}\BibitemShut {NoStop}%
	\bibitem [{\citenamefont {Lerosey}\ \emph {et~al.}(2007)\citenamefont
		{Lerosey}, \citenamefont {Rosny}, \citenamefont {Tourin},\ and\ \citenamefont
		{Fink}}]{lerosey_focusing_2007_1}%
	\BibitemOpen
	\bibfield  {author} {\bibinfo {author} {\bibfnamefont {G.}~\bibnamefont
			{Lerosey}}, \bibinfo {author} {\bibfnamefont {J.~d.}\ \bibnamefont {Rosny}},
		\bibinfo {author} {\bibfnamefont {A.}~\bibnamefont {Tourin}},\ and\ \bibinfo
		{author} {\bibfnamefont {M.}~\bibnamefont {Fink}},\ }\href
	{https://doi.org/10.1126/science.1134824} {\bibfield  {journal} {\bibinfo
			{journal} {Science}\ }\textbf {\bibinfo {volume} {315}},\ \bibinfo {pages}
		{1120} (\bibinfo {year} {2007})}\BibitemShut {NoStop}%
	\bibitem [{\citenamefont {Xu}\ \emph {et~al.}(2011)\citenamefont {Xu},
		\citenamefont {Liu},\ and\ \citenamefont {Wang}}]{xu_time-reversed_2011_1}%
	\BibitemOpen
	\bibfield  {author} {\bibinfo {author} {\bibfnamefont {X.}~\bibnamefont
			{Xu}}, \bibinfo {author} {\bibfnamefont {H.}~\bibnamefont {Liu}},\ and\
		\bibinfo {author} {\bibfnamefont {L.~V.}\ \bibnamefont {Wang}},\ }\href
	{https://doi.org/10.1038/nphoton.2010.306} {\bibfield  {journal} {\bibinfo
			{journal} {Nat. Photonics}\ }\textbf {\bibinfo {volume} {5}},\ \bibinfo
		{pages} {154} (\bibinfo {year} {2011})}\BibitemShut {NoStop}%
	\bibitem [{\citenamefont {Judkewitz}\ \emph {et~al.}(2013)\citenamefont
		{Judkewitz}, \citenamefont {Wang}, \citenamefont {Horstmeyer}, \citenamefont
		{Mathy},\ and\ \citenamefont {Yang}}]{judkewitz_speckle-scale_2013_1}%
	\BibitemOpen
	\bibfield  {author} {\bibinfo {author} {\bibfnamefont {B.}~\bibnamefont
			{Judkewitz}}, \bibinfo {author} {\bibfnamefont {Y.~M.}\ \bibnamefont {Wang}},
		\bibinfo {author} {\bibfnamefont {R.}~\bibnamefont {Horstmeyer}}, \bibinfo
		{author} {\bibfnamefont {A.}~\bibnamefont {Mathy}},\ and\ \bibinfo {author}
		{\bibfnamefont {C.}~\bibnamefont {Yang}},\ }\href
	{https://doi.org/10.1038/nphoton.2013.31} {\bibfield  {journal} {\bibinfo
			{journal} {Nat. Photonics}\ }\textbf {\bibinfo {volume} {7}},\ \bibinfo
		{pages} {300} (\bibinfo {year} {2013})}\BibitemShut {NoStop}%
	\bibitem [{\citenamefont {Bosch}\ \emph {et~al.}(2016)\citenamefont {Bosch},
		\citenamefont {Goorden},\ and\ \citenamefont {Mosk}}]{bosch_frequency_2016_1}%
	\BibitemOpen
	\bibfield  {author} {\bibinfo {author} {\bibfnamefont {J.}~\bibnamefont
			{Bosch}}, \bibinfo {author} {\bibfnamefont {S.~A.}\ \bibnamefont {Goorden}},\
		and\ \bibinfo {author} {\bibfnamefont {A.~P.}\ \bibnamefont {Mosk}},\ }\href
	{https://doi.org/10.1364/OE.24.026472} {\bibfield  {journal} {\bibinfo
			{journal} {Opt. Express}\ }\textbf {\bibinfo {volume} {24}},\ \bibinfo
		{pages} {26472} (\bibinfo {year} {2016})}\BibitemShut {NoStop}%
	\bibitem [{\citenamefont {Zhou}\ \emph {et~al.}(2014)\citenamefont {Zhou},
		\citenamefont {Ruan}, \citenamefont {Yang},\ and\ \citenamefont
		{Judkewitz}}]{zhou_focusing_2014_1}%
	\BibitemOpen
	\bibfield  {author} {\bibinfo {author} {\bibfnamefont {E.~H.}\ \bibnamefont
			{Zhou}}, \bibinfo {author} {\bibfnamefont {H.}~\bibnamefont {Ruan}}, \bibinfo
		{author} {\bibfnamefont {C.}~\bibnamefont {Yang}},\ and\ \bibinfo {author}
		{\bibfnamefont {B.}~\bibnamefont {Judkewitz}},\ }\href
	{https://doi.org/10.1364/OPTICA.1.000227} {\bibfield  {journal} {\bibinfo
			{journal} {Optica}\ }\textbf {\bibinfo {volume} {1}},\ \bibinfo {pages} {227}
		(\bibinfo {year} {2014})}\BibitemShut {NoStop}%
	\bibitem [{\citenamefont {Ma}\ \emph {et~al.}(2014)\citenamefont {Ma},
		\citenamefont {Xu}, \citenamefont {Liu},\ and\ \citenamefont
		{Wang}}]{ma_time-reversed_2014_1}%
	\BibitemOpen
	\bibfield  {author} {\bibinfo {author} {\bibfnamefont {C.}~\bibnamefont
			{Ma}}, \bibinfo {author} {\bibfnamefont {X.}~\bibnamefont {Xu}}, \bibinfo
		{author} {\bibfnamefont {Y.}~\bibnamefont {Liu}},\ and\ \bibinfo {author}
		{\bibfnamefont {L.~V.}\ \bibnamefont {Wang}},\ }\href
	{https://doi.org/10.1038/nphoton.2014.251} {\bibfield  {journal} {\bibinfo
			{journal} {Nat. Photonics}\ }\textbf {\bibinfo {volume} {8}},\ \bibinfo
		{pages} {931} (\bibinfo {year} {2014})}\BibitemShut {NoStop}%
	\bibitem [{\citenamefont {Ruan}\ \emph {et~al.}(2017)\citenamefont {Ruan},
		\citenamefont {Haber}, \citenamefont {Liu}, \citenamefont {Brake},
		\citenamefont {Kim}, \citenamefont {Berlin},\ and\ \citenamefont
		{Yang}}]{ruan_focusing_2017_1}%
	\BibitemOpen
	\bibfield  {author} {\bibinfo {author} {\bibfnamefont {H.}~\bibnamefont
			{Ruan}}, \bibinfo {author} {\bibfnamefont {T.}~\bibnamefont {Haber}},
		\bibinfo {author} {\bibfnamefont {Y.}~\bibnamefont {Liu}}, \bibinfo {author}
		{\bibfnamefont {J.}~\bibnamefont {Brake}}, \bibinfo {author} {\bibfnamefont
			{J.}~\bibnamefont {Kim}}, \bibinfo {author} {\bibfnamefont {J.~M.}\
			\bibnamefont {Berlin}},\ and\ \bibinfo {author} {\bibfnamefont
			{C.}~\bibnamefont {Yang}},\ }\href {https://doi.org/10.1364/OPTICA.4.001337}
	{\bibfield  {journal} {\bibinfo  {journal} {Optica}\ }\textbf {\bibinfo
			{volume} {4}},\ \bibinfo {pages} {1337} (\bibinfo {year} {2017})}\BibitemShut
	{NoStop}%
	\bibitem [{\citenamefont {Fan}\ and\ \citenamefont
		{Kahn}(2005)}]{fan_principal_2005_1}%
	\BibitemOpen
	\bibfield  {author} {\bibinfo {author} {\bibfnamefont {S.}~\bibnamefont
			{Fan}}\ and\ \bibinfo {author} {\bibfnamefont {J.~M.}\ \bibnamefont {Kahn}},\
	}\href {https://doi.org/10.1364/OL.30.000135} {\bibfield  {journal} {\bibinfo
			{journal} {Opt. Lett.}\ }\textbf {\bibinfo {volume} {30}},\ \bibinfo {pages}
		{135} (\bibinfo {year} {2005})}\BibitemShut {NoStop}%
	\bibitem [{\citenamefont {Carpenter}\ \emph {et~al.}(2015)\citenamefont
		{Carpenter}, \citenamefont {Eggleton},\ and\ \citenamefont
		{Schröder}}]{carpenter_observation_2015_1}%
	\BibitemOpen
	\bibfield  {author} {\bibinfo {author} {\bibfnamefont {J.}~\bibnamefont
			{Carpenter}}, \bibinfo {author} {\bibfnamefont {B.~J.}\ \bibnamefont
			{Eggleton}},\ and\ \bibinfo {author} {\bibfnamefont {J.}~\bibnamefont
			{Schröder}},\ }\href {https://doi.org/10.1038/nphoton.2015.188} {\bibfield
		{journal} {\bibinfo  {journal} {Nat. Photonics}\ }\textbf {\bibinfo {volume}
			{9}},\ \bibinfo {pages} {751} (\bibinfo {year} {2015})}\BibitemShut {NoStop}%
	\bibitem [{\citenamefont {Xiong}\ \emph {et~al.}(2016)\citenamefont {Xiong},
		\citenamefont {Ambichl}, \citenamefont {Bromberg}, \citenamefont {Redding},
		\citenamefont {Rotter},\ and\ \citenamefont
		{Cao}}]{xiong_spatiotemporal_2016_1}%
	\BibitemOpen
	\bibfield  {author} {\bibinfo {author} {\bibfnamefont {W.}~\bibnamefont
			{Xiong}}, \bibinfo {author} {\bibfnamefont {P.}~\bibnamefont {Ambichl}},
		\bibinfo {author} {\bibfnamefont {Y.}~\bibnamefont {Bromberg}}, \bibinfo
		{author} {\bibfnamefont {B.}~\bibnamefont {Redding}}, \bibinfo {author}
		{\bibfnamefont {S.}~\bibnamefont {Rotter}},\ and\ \bibinfo {author}
		{\bibfnamefont {H.}~\bibnamefont {Cao}},\ }\href
	{https://doi.org/10.1103/PhysRevLett.117.053901} {\bibfield  {journal}
		{\bibinfo  {journal} {Phys. Rev. Lett.}\ }\textbf {\bibinfo {volume} {117}},\
		\bibinfo {pages} {053901} (\bibinfo {year} {2016})}\BibitemShut {NoStop}%
\end{thebibliography}

\end{document}